\documentclass[a4paper,fleqn,usenatbib]{aa}

\usepackage[varg]{txfonts}
\usepackage{graphicx}	
\usepackage{amsmath}	
\usepackage{amssymb}	
\usepackage{multicol}
\usepackage{bm}		
\usepackage{natbib,hyperref,ifthen} 
\usepackage{hyperref}
\usepackage{etoolbox}
\hypersetup{
 colorlinks=true,
 citecolor=blue}
\usepackage{color}   
\newcommand{\kms}{\textrm{km s}^{-1}} 
 \renewcommand\appendix{\par
  \setcounter{section}{0}
  \setcounter{subsection}{0}
  \setcounter{figure}{0}
  \setcounter{table}{0}
  \renewcommand\thesection{ \Alph{section}}
  \renewcommand\thefigure{\Alph{section}\arabic{figure}}
  \renewcommand\thetable{\Alph{section}\arabic{table}}
}

\usepackage[T1]{fontenc}
\usepackage{ae,aecompl}

\begin{document}

   \title{Simulating non-axisymmetric flows in disc galaxies}
   \titlerunning{Simulating non-axisymmetric flows in disc galaxies}

   \author{T. H. Randriamampandry
          \inst{1,2}
          \and
          N. Deg\inst{1}
          \and
          C. Carignan\inst{1,3}
          \and
           L. M. Widrow\inst{4}
            }

   \institute{Department of Astronomy, University of Cape Town, Private Bag X3, Rondebosch 7701, South Africa
            \and
            Kavli Institute for Astronomy and Astrophysics, Peking University, Beijing 100871, China \\
            \email{rtoky@pku.edu.cn}
         \and
             Observatoire d'Astrophysique de l'Universit\'e de Ouagadougou (ODAUO), BP 7021, Ouagadougou 03, Burkina Faso           
           \and
             Department of Physics, Engineering Physics, and Astronomy, Queen's University, Kingston, ON, K7L 3N6, Canada
             }

   \date{Received ....; accepted ....}

  \abstract
   {We present a two-step method to simulate and study non-circular motions in strongly barred galaxies. The first step is to constrain the initial parameters using a Bayesian analysis of each galaxy's azimuthally averaged rotation curve, the 3.6 $\rm \mu m$ surface brightness, and the gas surface density.  The second step is to generate equilibrium models using the GalactICS code and evolve them via GADGET-2.}
   {The bar strengths and mock velocity maps of the resulting snapshots are compared to observations in order to determine the best representation of the galaxy. }
   {We test our method on the unbarred galaxy NGC 3621 and the barred galaxies NGC 1300 and NGC 1530. NGC 3621 provides a validation of our method of generating initial conditions. NGC 1530 has an intermediate bar orientation that allows for a comparison to DiskFit. Finally NGC 1300 has a bar oriented parallel to the galaxy's major axis, where other algorithms tend to fail. }
   {Our models for NGC 3621 and NGC 1530 are comparable to those obtained using  commonly available algorithms.  Moreover, we have produced one of the first mass distribution models for NGC 1300. }
   {}


   \maketitle
   
\hfill
\section{Introduction}
Galaxy rotation curves (RCs) that are derived from spectroscopic observations of the gas content in 
disc galaxies can be used to infer the distribution of dark matter on galactic scales
(e.g. \citealt{de-Blok:2008oq}).   This inference relies on the assumption that the 
gas moves along circular orbits.  However, in barred galaxies, non-circular flows in the gas disc 
lead to systematic errors in the measured RCs (e.g. \citealt{2008MNRAS.385..553D}). 
Investigations like those of \citet{1978PhDT.......195B, 1981AJ.....86.1825B} 
have determined that the kinematics of barred galaxies are non-axisymmetric and that the bar produces oval or bi-symmetric distortions (e.g. \citealt{1977A&A....57...97B,1977A&A....57..373B,1980A&A....90..123V, 1996A&A...313...65L,2007ApJ...664..204S,2008AJ....136.2761O}).

Two of the most well-known methods used when deriving RCs from velocity maps of a galaxy's gas content are ROTCUR \citep{1989A&A...223...47B} and 
DiskFit \citep{2007ApJ...664..204S}.  ROTCUR is an implementation of the tilted-ring method where the gas is assumed to be on circular orbits \citep{1974ApJ...193..309R}.  As such, it under/over-estimates the RC when the bar is parallel/perpendicular to the major axis of the observed galaxy (e.g. \citealt{2015MNRAS.454.3743R}). 
Quantitatively, systematic errors of the rotational velocities can be up to 40\% of the 
 expected values when the bar is within $10^{\circ}$ of the major or minor axis of 
 the galaxy \citep{2016A&A...594A..86R}. 
 DiskFit is specifically designed to account for the non-circular motions mainly induced by 
 bars \citep{2007ApJ...664..204S}. It works quite well for bars with an orientation intermediate 
 to the major and minor axes. However, when the bar orientation approaches the major or minor axis, a 
 degeneracy in the fitting equation causes the algorithm to fail \citep{2010MNRAS.404.1733S}. 
 
 An alternative method is to use hydrodynamical simulations to understand and quantify the 
non-circular motions produced by a bar (e.g. \citealt{1999ApJ...522..699A,2002Ap&SS.281...39A,2002MNRAS.330...35A,2013MNRAS.429.1949A}). Bars arise naturally in many hydrodynamical simulations (e.g. \citealt{1991A&A...252...75P,1991A&A...243..118S}), allowing for a deep exploration of bar dynamics. These simulations can be used to understand non-circular motions in two distinct ways.  Firstly, snapshots from arbitrary barred galaxy simulations can be observed and analyzed using algorithms like ROTCUR and DiskFit (e.g. \citealt{1983ApJ...275..529K,1990ApJ...357...71O,2000A&A...362..435L, 2012MNRAS.427.2523K, 2016A&A...594A..86R}).  These pseudo-observations are then compared to the gravitational potential to understand the systematic errors and biases introduced by the bar's non-circular motions (see \citealt{2016A&A...594A..86R}).
The other approach is to model specific galaxies using simulations. This is done by simulating a number of plausible galaxy initial conditions.  Then the snapshots from each simulation are observed at the same orientation/inclination as the target galaxy.  The snapshot that provides the best match to the data is used to infer the galaxy's mass.
\citet{1983A&A...121..297D} attempted to model NGC 5383 using this methodology.  They used the axisymmetric RCs and surface brightness profile to fix the general galaxy model.  However, they artificially grow a bar rather than allowing it to develop dynamically. This method has been also used by \citet{1986MNRAS.221....1T} and later by \citet{2008ApJ...674..797Z} for NGC 1365 and \citet{1988A&A...189...59B} for NGC 6946.


In this work we present a new method to model barred spiral galaxies based on Bayesian statistics and hydrodynamical simulations.
 Firstly a Bayesian analysis on the azimuthally-averaged RC, gas surface density, and 3.6 $\mu m$  surface brightness is performed to constrain the general galaxy parameters. This  step is similar to the MagRite method of modeling SAMI IFU data \citep{tat2017}, where the main difference is that they use 2D IFU data and we use H{\sc i} data. Next, a set of models that vary in their susceptibility to bar formation is selected from the Bayesian analysis. Initial conditions are drawn from the multi-component, axisymmetric equilibrium models as generated by the code GalactICs \citep{Widrow2008}.  
 
 For a particular galaxy we determine the Probability Distribution Functions (PDF) in the Baysian parameter space by comparing observational data  with model  predictions. We then select three examples from the PDF to be used as initial conditions for the hydrodynamical N-body simulations. The galaxies are allowed to evolved for 5 Gyr. The snapshot that best reproduces the galaxy is selected by comparing the maximum bar strength measured from the simulation with the observed bar strength for each galaxy. We then make mock observations of the selected snapshot using the orientation parameters for each galaxy, which are compared with the true ones.  
 
This methodology is fundamentally different from the \citet{1983A&A...121..297D} algorithm in two key ways.  Firstly, the Bayesian analysis explores a significantly larger space of possible models.  Secondly, the bars grow and buckle self-consistently rather than being grown from a static, arbitrarily imposed potential.

In Section \ref{sec:options} we present the observations of NGC 3621, 1530, and 1300.
Section \ref{sec:methods} describes the GalactICs algorithm.  Section \ref{meth} explains the methodology. In Section \ref{sec:Comparison} we present and discuss the results. Finally, Section \ref{sec:summary} contains our conclusions.

\section{Data from observations}
\label{sec:options}
In this section, we present the galaxies in our sample and 
the data that are used to constrain the initial conditions. These 
observations comprise: the RC, the gas density profile and 3.6$\rm \mu m$
 surface brightness profile.

The sample galaxies were selected depending on their bar properties and the availability of H{\sc i} and near-infrared observations. 
Table \ref{tab:1} summarizes their properties. 
The sample includes two strongly barred galaxy NGC 1300 and NGC 1530 and the late-type unbarred spiral galaxy NGC 3621.

\begin{itemize}
\item NGC 3621 is included in the sample to test the basic methodology. It is a late type unbarred spiral galaxy located at a  distance of 6.6 Mpc \citep{2001ApJ...553...47F}. This galaxy is  part of The H{\sc i} Nearby  Galaxies Survey (THINGS, \citealt{2008AJ....136.2563W}). THINGS data products are publicly available and can be obtained from their webpage (\url{http://www.mpia.de/THINGS/Data.html}).  The velocity map is derived from the natural weighted datacube using the GIPSY task MOMENTS. The RC is then derived from the velocity map using the GIPSY task ROTCUR. The gas density profile is obtained from the moment0 map
 using the ELLINT task and assuming the kinematic parameters from \citet{de-Blok:2008oq}. 
 The 3.6 $\rm \mu m$ surface brightness profiles are obtained from de Blok (private communication).
 
\item NGC 1530 is a barred spiral galaxy, SBb, located at a distance of 19.9 Mpc  \citep{2014MNRAS.444..527S}. This galaxy was chosen to compare our method with DiskFit. The H{\sc i} observations were obtained using the DnC array configuration of the VLA. The raw data were retrieved from the archive, edited and calibrated using the standard NRAO Astronomical Image Processing System (AIPS) packages. Data cubes were then obtained and cleaned with the AIPS task {\sc imagr} using robust0 weighting, which optimizes sensitivity and spatial resolution. 0237-2330 is the phase and amplitude calibrator and 137+331 the flux calibrator. The cleaned data cube has 64 channels with 10.43 km s$^{-1}$ velocity resolution and 14 arcsec angular resolution, corresponding to a linear physical resolution of $\sim$1.3 kpc at the adopted distance of 19.9 Mpc.
This galaxy has a strong bar with intermediate orientation, $\Phi_{b}$ = -25$^{\circ}$ \citep{2007ApJ...657..790M}. The 3.6 $\rm \mu m$ surface brightness profile was obtained from SPITZER images retrieved from the archive and derived using the IRAF task {\sc ellipse}. The image was first cleaned from foreground stars before being used in the {\sc ellipse} task. The profile was then corrected for inclination.

\item NGC 1300 is a grand design barred spiral galaxy, SBbc, located at a distance of 17.1 Mpc \citep{1981ApJ...248..395D}. The bar is almost perfectly aligned with the major axis \citep{2000A&A...361..841A}. As such, this galaxy cannot be analyzed using ROTCUR or DiskFit or most other standard methodologies. The H{\sc i} observations are obtained using the VLA C configuration. The data were firstly analyzed by \citet{1989ApJ...337..191E} and later by \citet{1997A&A...317...36L}, which reported different values for the inclination and position angle of the disc. We re-analyzed the data and found the same kinematic parameters as those reported by \citet{1997A&A...317...36L} (listed in Table \ref{tab:1}). For this re-analysis the raw data were retrieved from the archive, edited and calibrated using AIPS. Data cubes were then obtained and cleaned with the AIPS task {\sc imagr} using robust0 weighting. 1031+561 is the phase and amplitude calibrator and 3C286 the flux calibrator. The resulting data cube has 64 channels with 20.8  km s$^{-1}$ velocity resolution and 20 arcsec angular resolution, which corresponds to a linear physical resolution of $\sim$1.6 kpc at the adopted distance of 17.1 Mpc. 
The orientation parameters listed in Table 1 were then used to derive the RC and the gas density profile. The 3.6 $\rm \mu m$ surface brightness profile was derived using the same method as for NGC1530.
\end{itemize}

\begin{table*}
 \caption{Summary of the properties of the galaxies in the sample.}
 \label{tab:1}
 \scriptsize
 \begin{center}
 \begin{tabular}{lllllllll}
  \hline\\
 Name &V$_{sys}$ (km/s)&Morph. type&D (Mpc) &  P. A. ($\circ$) & Incl. ($\circ$) & Bar strength (Q$_{b}$)&$\phi_{bar}$($\circ$)&R$_{bar}$(arcsec)\\\\
  1 &2&3&4 & 5& 6 & 7&8&9\\\\
  \hline\\
  NGC3621&729[dB08]&SAd&6.6[F01]&345[dB08]&65[dB08]&-&-&-\\\\[2pt]
NGC1530&2461[Z04]&SBb&19.9[S14]&188[Z04]&53[Z04] &0.39[A98]&-25 [MD07]&37[A98]\\\\[2pt]
NGC1300&1575[Lin97]&SBbc&17.1[dV81]&267[Lin97]&35[Lin97]&0.537[L04]&-2 [L04]&87[L04]\\\\[2pt]
 \hline
\multicolumn{9}{l}{\textsuperscript{}\footnotesize{Notes}}\\
\multicolumn{9}{l}{\textsuperscript{}\footnotesize{A98: \citet{1998AJ....116.2136A}, dB08: \citet{de-Blok:2008oq}, L04: \citet{2004ApJ...607..103L}, }}\\
\multicolumn{9}{l}{\textsuperscript{}\footnotesize{Lin97: \citet{1997A&A...317...36L}, MD07: \citet{2007ApJ...657..790M},FO1:\cite{2001ApJ...553...47F}}}\\
\multicolumn{9}{l}{\textsuperscript{}\footnotesize{S14: \cite{2014MNRAS.444..527S}, dV81:\cite{1981ApJ...248..395D}, Z04:\cite{2004A&A...413...73Z}.}}
 \end{tabular}
   \end{center}
\end{table*}
\section{GalactICs }
\label{sec:methods}



We build our model galaxies using a new version of GalactICS \citep{1995MNRAS.277.1341K,Widrow2008,Deg2018} that includes a gas disc component. 
In this work, we generate galaxies with a S\'ersic bulge, an exponential stellar disc, an exponential gas disc, and a double-power law dark matter halo.

The density profile of the S\'ersic bulge is \citep{2010ApJ...715.1152P} 
\begin{equation}
\rho_{\rm bulge}(r)= \rho_{b}\left(\frac{r}{r_{b}}\right)^{-p} e^{-b\left(r/r_{b}\right)^{1/n}},
\end{equation}
where $r$ is the spherical radius and $p = 1 - 0.6097/n +
0.05563/n^{2}$ yields the S\'ersic profile \citep{1997A&A...321..111P,2005MNRAS.362..197T}.
For simplicity, the bulge is characterized by a scale velocity, $\sigma_{b}$, given 
by
\begin{equation}
 \sigma_{b}= [4\pi n b^{n(p-2)}(n(2-p))r^{2}_{b}\rho_{b}]^{1/2}.
\end{equation}
GalactICs assumes an exponential-sech$^{2}$ \citep{1981A&A....95..105V} density distribution for the disc:
 \begin{equation}
 \rho_{d}(R,z)= \frac{M_{d}}{4 \pi R_{d}^{2} z_{d}} 
e^{(-R/R_{d})}\textrm{sech}^{2}(z/z_{d}),
\end{equation}
where $M_{d}$ is the disc mass, $R_{d}$ is the disc scale length, and z$_{d}$ is the disc scale height and R is the cylindrical radius. 
The radial velocity dispersion is given by
\begin{equation}
 \sigma_{R}^{2}(R) = \sigma_{0}^{2}e^{-(R/R_{\sigma})},
\label{eq4}
\end{equation}
 where $\sigma_{0}$ and R$_{\sigma}$ are the central dispersion and a scale radius respectively. Following \citet{Widrow2008}, $R_{\sigma}$ is set equal to the scale length based on the observations of  \cite{1993A&A...275...16B}. 


 However, in terms of the stability, the key value is the Toomre Q parameter calculated at R = 2.2 R$_{d}$. Given the use of R$_{\sigma}$ = R$_{d}$, the variation in $\sigma_{0}$ gives only small variations in Q, and therefore of the disk stability. Therefore, for simplicity we fixed $\sigma_{0}$ to 90 km s$^{-1}$.
The disk is also truncated at R$_{t}$= 50 kpc for all models.

The gas disc has an exponential surface density. Following \citet{2010MNRAS.407..705W}, the disc
 is initially isothermal, which requires a scale height that increases with radius.
 
The double power-law halo is defined as (see \citealt{Widrow2008})
\begin{equation}
 \rho(R) = \frac{2^{2-\alpha}\sigma_{h}^{2}}{4 \pi r_{h}^{2}}
\frac{1}{u^{\alpha}(1+u)^{\beta-\alpha}}C(r;r_{t},\delta r_{t})~,
\end{equation}
where $u=r/r_{h}$, $\sigma_{h}$ is a scale velocity, and $r_{h}$ is the scale radius.  The 
$C(r;r_{t},\delta r_{t})$ term is a truncation factor, given by
\begin{equation}
  C(r;r_{t},\delta r_{t})=\frac{1}{2}\textrm{erfc}\left(\frac{r-r_{t}}{\sqrt{2}\delta r_{h}} \right).
\end{equation}
In this work the truncation parameters are set to $r_{t}=$ 200 kpc and $\delta r_{t}$=20 kpc

 \begin{table*}
 \small
 \caption{Summary of the model parameters.}
 \label{tab:2}
 \begin{center}
 \begin{tabular}{llllllllll}
  \hline\\
 Parameter &Units &NGC3621& & &NGC1530&   & &NGC1300& \\\\
 \hline
  & & Min.&Max.&& Min. &Max. &&Min.&Max.\\\\
 1 &2 &3 &4  &   &5& 6 & & 7&8 \\
  \hline\\
Halo scale velocity: $\sigma_{h}$ & km s$^{-1}$ & 10&400&&10&400&&10&400\\[2pt]
Halo scale radius: $R_{h}$& kpc&5.0&20.0&&2.0&20.0 &&1.0&20.0\\[2pt]
 Inner slope:$\alpha$& - &0.1 &2.0&&0.1&2.0&&0.1&2.0\\[2pt]
 Outer slope:$\beta$ & - & 1.5&4.0&&1.5&4.0&&1.5&4.0\\[2pt]
 Stellar disc
 mass:$M_{d}$ & 10$^{10} \ M_{\sun}$ &0.08 &11.8&&0.05&10.5&&0.05&16.5\\[2pt]
 Disc scale length: $R_{d}$ & kpc & 1.0&20.0&&1.0&6.5&&0.5&6.5\\[2pt]
 Gas mass:$M_{g}$ & 10$^{10} \ M_{\sun}$ &0.02&3.5&&0.01&3.5&&0.02&3.5\\[2pt]
 Gas scale length: $R_{g}$ & kpc & 1.0&20.0&&0.5&20.0&&1.0&20.0\\[2pt]
Bulge S\'ersic index: n& - &0.01&5.0&&0.01&5.0&&0.01&5.0\\[2pt]
  Characteristic bulge velocity:$\sigma_{b}$& km s$^{-1}$ &10&500&&10&250&&10&600\\[2pt]
Bulge scale length:$R_{b}$ & kpc &0.1&2.5&&0.05&2.5&&0.05&2.5\\[2pt]
  Disk mass-to-light ratio:$\Upsilon_{D}$& $M_{\sun}/L_{\sun}$&0.01&4.00&&0.01&4.00&&0.01&4.00\\[2pt]
Bulge mass-to-light ratio:$\Upsilon_{B}$ &  $M_{\sun}/L_{\sun}$&0.01&4.00&&0.01&4.00&&0.01&4.00\\\\[2pt]
  \hline
\multicolumn{7}{l}{\footnotesize{Notes}}\\
\multicolumn{7}{l}{\footnotesize{ The parameters' name are in column 1, the units are given in column 2.}}\\
\multicolumn{7}{l}{\footnotesize{The lower and upper bounds are shown in columns 3, 5, 7 and 4, 6, 8 respectively.}}\\
 \end{tabular}
   \end{center}
\end{table*}

 \begin{table*}
 \caption{Comparison between the model input parameters and the THINGS mass model results for NGC 3621. }
 \label{tab:3}
 \begin{center}
 \begin{tabular}{llllllll}
  \hline\\
 Parameter &Units&Best fit&Model-A&Model-B&Model-C&THINGS-ISO&THINGS-NFW \\\\
 \hline\\
 1 &2&3&4 &5 &6&7&8    \\
  \hline\\
 $\sigma_{h}$ &km s$^{-1}$&275 $\pm$ 32& 280&275 &265&-&-\\[2pt]
$R_{h}$ &kpc&15.26 $\pm$ 2.08& 14.50&15.26&16.01&-&-\\[2pt]
 $\alpha$&-&0.52 $\pm$ 0.22& 0.50 &0.52 &0.12&0&1\\[2pt]
 $\beta$ &-&2.80 $\pm$ 0.32& 3.12& 2.80&2.75&2&3\\[2pt]
 $M_{d}$ &10$^{10} \ M_{\sun}$&1.90 $\pm$ 0.31& 2.20&1.90&1.38&1.95[dB08]&1.95[dB08]\\[2pt]
 $R_{d}$ &kpc&2.45 $\pm$ 0.25 & 2.60& 2.45&2.62&2.61[W08]&2.61[W08]\\[2pt]
 $M_{g}$ &10$^{10} \ M_{\sun}$&0.89 $\pm$ 0.05& 0.94 &0.89&1.11&0.70[W08]&0.70[W08]\\[2pt]
 $R_{g}$ &kpc&14.89 $\pm$ 0.03& 15.00&14.98&12.32&-&-\\[2pt]

  $\sigma_{b}$&km s$^{-1}$&62 $\pm$ 17&50 &62&40&-&-\\[2pt]
$R_{b}$ &kpc&0.96 $\pm$ 0.22& 0.60 &0.96&1.01&-&-\\[2pt]
$M_{b}$ &10$^{10} \ M_{\sun}$&0.17 $\pm$ 0.01&0.07 &0.17&0.08&-&-\\[2pt]
 $\rho_{0}$ &10$^{-3}$ M$_{\odot}$pc$^{-3}$&-&-&-&-&14.4[dB08]&-\\[2pt]
 $R_{c}$ &kpc&-&-&-&-&5.54[dB08]&-\\[2pt]
  $V_{200}$ &km s$^{-1}$&-&-&-&-&-&165[dB08]\\[2pt]
  $R_{s}$ &kpc&-&-&-&-&-&28.8[dB08]\\[2pt]
   $\Upsilon_{D}$ & $M_{\sun}/L_{\sun}$&0.31$\pm$0.05&0.31&0.31&0.31&0.60$^{a}$&0.60\\[2pt]
 $\Upsilon_{B}$ & $M_{\sun}/L_{\sun}$&1.13$\pm$0.35&1.13&1.13&1.13&-&-\\[2pt]
 $Q_{d}$ &-&-&0.53&0.49&0.34&-&-\\[2pt]
 $X_{d}$ &-&-&1.75&1.57&1.26&-&-\\[2pt]
  \hline
 \multicolumn{8}{l}{\footnotesize{Notes}}\\
\multicolumn{8}{l}{\footnotesize{ $\Upsilon_{D}$ and $\Upsilon_{D}$ are the best fit mass-to-light ratio for the stellar disk and bulge respectively}}\\
\multicolumn{8}{l}{\footnotesize{ $Q_{d}$ and $X_{d}$ are the disc stability parameters  measured at 2.2 R$_{d}$}}\\
 \multicolumn{8}{l}{\footnotesize{The ISO and NFW THINGS mass model results with fixed mass-to light ratio are shown in column 7 and 8 for comparison}}\\ 
\multicolumn{8}{l}{\footnotesize{dB08: \citet{de-Blok:2008oq}, W08: \citet{2008AJ....136.2563W}, $R_{c}$ and $R_{s}$ are the ISO and NFW scale radius respectively}}\\ 
\multicolumn{8}{l}{\footnotesize{$^{a}$ $\Upsilon_{*}$ as measured from a diet-Salpeter IMF}}\\
\multicolumn{8}{l}{\footnotesize{It is worth noting that $\Upsilon_{*}$ measured using the Kroupa IMF is 0.42  \citep{de-Blok:2008oq}.}}\\
 \end{tabular}
   \end{center}
\end{table*}

 \begin{table*}
 \caption{NGC1530 Model input parameters. }
 \label{tab:4}
 \begin{center}
 \begin{tabular}{llllll}
  \hline\\
 Parameter &Units&best fit&Model-A&Model-B&Model-C \\\\
 \hline\\
 1 &2&3&4 &5 &6    \\
  \hline\\
 $\sigma_{h}$ &km s$^{-1}$&310 $\pm$ 19&325&310 &310\\[2pt]
$R_{h}$ &kpc&16.79 $\pm$ 1.25&16.14&16.79&14.76\\[2pt]
 $\alpha$&-&0.34 $\pm$ 0.09& 0.33&0.34&0.24 \\[2pt]
 $\beta$ &-&2.42 $\pm$ 0.16&2.59&2.42&2.55 \\[2pt]
 $M_{d}$ &10$^{10} \ M_{\sun}$&3.53 $\pm$ 0.29& 3.00&3.53&3.16\\[2pt]
 $R_{d}$ &kpc&3.36 $\pm$ 0.29 & 3.88& 3.36&3.48\\[2pt]
 $M_{g}$ &10$^{10} \ M_{\sun}$&1.39 $\pm$ 0.02& 1.33 &1.39&1.33\\[2pt]
 $R_{g}$ &kpc&6.19 $\pm$ 0.07& 6.06&6.19&6.14\\[2pt]
 $\sigma_{b}$&km s$^{-1}$&44 $\pm$ 20&45 &44&49\\[2pt]
$R_{b}$ &kpc&0.65 $\pm$ 0.06& 0.72 &0.65&0.59\\[2pt]
 $M_{b}$ &10$^{10} \ M_{\sun}$&0.07 $\pm$ 0.01& 0.08&0.07&0.08\\[2pt]
$\Upsilon_{D}$ & $M_{\sun}/L_{\sun}$&1.09$\pm$0.07&1.09&1.09&1.09\\[2pt]
$\Upsilon_{B}$ & $M_{\sun}/L_{\sun}$&1.26$\pm$0.10&1.26&1.26&1.26\\[2pt]
$Q_{d}$ &-&-&0.70&0.56&0.38\\[2pt]
$X_{d}$ &-&-&3.38&3.11&2.40\\[2pt]
  \hline\\
 \end{tabular}
   \end{center}
\end{table*}

 \begin{table*}
 \caption{NGC1300 Model input parameters. }
 \label{tab:5}
 \begin{center}
 \begin{tabular}{llllll}
  \hline\\
 Parameter &Units&best fit&Model-A&Model-B&Models-C \\\\
 \hline\\
 1 &2&3&4 &5 &6    \\
  \hline\\
 $\sigma_{h}$ &km s$^{-1}$&159 $\pm$ 58&242&159&255\\[2pt]
$R_{h}$ &kpc&10.88 $\pm$ 3.31&13.45&10.88&13.65\\[2pt]
 $\alpha$&-&0.76$\pm$ 0.25&0.68&0.76&0.31\\[2pt]
 $\beta$ &-&2.98 $\pm$ 0.56& 3.54&2.98&3.42\\[2pt]
 $M_{d}$ &10$^{10} \ M_{\sun}$&8.72 $\pm$ 1.34& 7.37&8.72&8.82\\[2pt]
 $R_{d}$ &kpc&5.59 $\pm$ 0.18 & 5.79& 5.59&5.81\\[2pt]
 $M_{g}$ &10$^{10} \ M_{\sun}$&0.52 $\pm$ 0.07& 0.55 &0.52&0.54\\[2pt]
 $R_{g}$ &kpc&9.85 $\pm$ 0.85&9.71&9.85&10.68\\[2pt]
$\sigma_{b}$&km s$^{-1}$&53 $\pm$ 18&70.0 &53&43\\[2pt]
$R_{b}$ &kpc&2.59 $\pm$ 0.99& 1.17 &2.59&2.21\\[2pt]
$M_{b}$ &10$^{10} \ M_{\sun}$&0.34 $\pm$ 0.02& 0.26&0.34&0.19\\[2pt]
 $\Upsilon_{D}$ & $M_{\sun}/L_{\sun}$&0.64$\pm$0.12&0.64&0.64&0.64\\[2pt]
 $\Upsilon_{B}$ & $M_{\sun}/L_{\sun}$&1.94$\pm$0.78&1.94&1.94&1.94\\[2pt]
 $Q_{d}$ &-&-&0.38&0.28&0.24\\[2pt]
 $X_{d}$ &-&-&1.45&1.15&1.02\\[2pt]
  \hline
 \end{tabular}
   \end{center}
\end{table*}

\section{Methodology}
\label{meth}
In this section, we outline the steps adopted to construct the models. The estimation method for the input parameters is outlined in Section \ref{ipe} and details about the simulations are given in Section \ref{snapshot}.
 
 \subsection{Model parameters}
\label{ipe}
 In order to find plausible initial conditions for the simulations, we first run a Bayesian analysis for each galaxy.  For this analysis we utilize analytic approximations of axisymmetric GalactICs models.  In this approximation, the gas disc is given a constant scale height, rather than the flaring profile built into GalactICs.
The models are constrained by the RC, the azimuthally averaged 3.6 $\rm \mu m$ stellar surface brightness, and the gas surface density.  In order to avoid any systematic errors caused by the presence of the bars in NGC 1530 and NGC 1300, we fit only the portion of the RC beyond two bar lengths, R$_{bar}$, based on the results of \citet{2016A&A...594A..86R}. We also exclude 3.6 $\rm \mu m$ data beyond the optical radius defined by $\rm R_{25}$.  Beyond this 
radius, contamination from the sky background as well as foreground sources affects the 
surface brightness profile, flattening its slope.

The log-likelihood for a particular model is

\begin{equation}
  \mathcal{L}(D|\Theta)=  - \frac{1}{2}\sum^{N}_{i=1}\left[ \ln\left(2\pi\sigma_{D,i}^{2}\right) + \left(\frac{(D_{i} - M(\Theta)_{i})^{2}}{\sigma^{2}_{D,i}}\right)\right]~,
\label{equa7}
\end{equation}
where $\Theta$ are the model parameters, $D_{i}$ are the observations, $\sigma_{D,i}$ are the uncertainties in the observations, $M(\Theta)_{i}$ are the mock observations from the model, and N is the number of observed data points.  The posterior probability for a model is
\begin{equation}
 p(\Theta|D,I) = \frac{p(\Theta|I) \mathcal{L}(D|\Theta)}{Z}, 
\end{equation}
where $p(\Theta|I)$ is the prior probability for a model and $Z$ is the evidence.
Since we are performing a parameter estimate, the evidence is simply a normalization 
factor that is canceled out in the analysis.

To explore the parameter space we use the EMCEE algorithm \citep{2013PASP..125..306F}.  
EMCEE is a parallel, affine-invariant Markov Chain Monte Carlo (MCMC) algorithm.  It operates 
by initializing an ensemble of walkers randomly throughout the parameter space.  
The walkers move throughout the space by first proposing a new position using 
stretches along the vector to another randomly selected walker.  The posterior of the 
proposal is evaluated and compared to the current posterior in order to determine 
whether the walker will move or not.  In this work, we use 200 walkers for 3500 steps, 
giving a total of $7\times 10^5$ likelihood calls.  For simplicity, the priors for 
each parameter are uniform. A list of the relevant parameters as well as their minimum and maximum values considered are given in Table \ref{tab:2}. 

The disc stability can be quantified using the Toomre $Q$ \citep{1964ApJ...139.1217T} and $X$  \citep{1978ApJ...222..850G,1979ApJ...233..857G} parameters. 

The Toomre $Q$ parameter is given by
\begin{equation}
 Q_{d} = \frac{\sigma_{r}\kappa}{3.36G\Sigma_{d}}~,
\end{equation}
for a stellar disc
 where $\sigma_{r}$ is the radial velocity dispersion at a given radius, 
$\kappa$ is the epicyclic frequency, $G$ is the gravitational constant and $\Sigma_{d}$ is the stellar surface density.

For the gas disc
 $Q_{g}$ is
\begin{equation}
 Q_{g} = \frac{c_{s}\kappa}{\pi G\Sigma_{g}}~,
\end{equation}
where c$_{s}$ is the sound speed of the gas and  $\Sigma_{g}$ the gas surface density \citep{2010MNRAS.407..705W}. 
It is important to note that Q is a local quantity that depends on the radius.  
The surface densities, $\Sigma$, velocity dispersion $\sigma_{r}$ (given by Equation \ref{eq4} ), 
sound speed $c_{s}$, and the epicyclic frequency $\kappa$ are all functions of the radius. 
The $Q$ parameters are measures of a disc's stability against local perturbations.  
For this work, we calculate $Q$ at $2.2~R_{d}$ as that is where the ratio of disk to halo gravity is maximal. 
Given our value of $\sigma_{0}=90 ~\kms$ and $\sigma_{r}(2.2R_{D})=10~\kms$ for all models.

The $X$ parameter is a measure of a disc's self-gravity and it indicates the disc's stability against 
global perturbations. It is given by
\begin{equation}
 X= \left(\frac{V_{t}}{V_{d}}\right)^{2}_{2.2R_{d}}~,
\end{equation}
where $V_{t}$ and $V_{d}$ are the total circular velocity and the circular velocity due 
to the disc respectively.  As with the $Q$ parameter, it is measured at $2.2 R_{d}$.
Unstable discs have large $X$ and low $Q$, while those with low $X$ and high $Q$ are significantly more stable. 

The formal errors obtained during the derivation of the gas surface density and the 3.6 $\rm \mu m$ surface brightness profiles do not reflect the true uncertainties in the measurements. In order to more accurately represent the uncertainties in these profiles we include two extra parameters; $\epsilon_{p, G}$ for the gas surface density and $\epsilon_{p, S}$  for the stellar surface brightness profiles. These are added in quadrature to the formal errors during the Bayesian analysis.
 For those profiles we set:
\begin{equation}
 \sigma_{G,i}^{2} = \epsilon_{G,i}^{2} + \epsilon_{p,G,i}^{2}~,
\end{equation}
for the gas density and 
\begin{equation}
 \sigma_{S,i}^{2} = \epsilon_{S,i}^{2} + \epsilon_{p,S,i }^{2}~,
\end{equation}
for the stellar surface brightness profile,
where $\epsilon_{G}$, $\epsilon_{S}$ are the measured uncertainty and $\epsilon_{p,G}$, $\epsilon_{p,G}$ are the error parameters.

The input parameters for the initial condition are selected based on the parameter Probability Distribution Functions (PDFs) and the disc stability parameters PDFs. We have chosen three models from the PDF for simulation. Model A has a low disk stability, MOdel B uses the peak of the parameter PDFs, and MOdel C has a high degree of stability. For these simulations, we select the snapshot that best reproduces the observed bar strength and velocity field.


\subsection{Numerical simulations}
\label{snapshot}

Armed with the GalactICs parameters for each galaxy, we generated our N-body realizations with 10$^{6}$ particles; $5\times10^{5}$ for the halo, 
$2\times10^{5}$ for the stellar disc, another $2\times10^{5}$ for the gas disc and $10^{5}$ for the bulge.  
The models were evolved for 5 Gyr using the GADGET-2 code with a softening length of 50 pc.  Snapshots were taken every 50 Myr in order to follow the evolution of the galaxy in detail.  A mock velocity field was calculated for each snapshot and compared to the velocity fields obtained for the actual galaxies.

In detail, the velocity fields for each galaxy are obtained using H{\sc i} observations of each galaxy 
combined with position angle, inclination, and systemic velocities listed in Table \ref{tab:1}.  The mock velocity fields are obtained by rotating and shifting the simulation to the same distance,
orientation, and systemic velocity as the actual galaxy and assuming a flat disc model with constant position and inclination angles for all the models. We generate mock velocity maps with the same pixel size and angular resolutions as the observed galaxies.  This allows for a pixel-by-pixel comparison of the model and observed velocity maps.  The snapshot that best represents each galaxy is selected based on both the bar strength and this velocity map comparison.

The bar strength of a galaxy can be defined by the amplitude of the Fourier  m=2 mode (see \citealt{2009A&A...495..491A,2016A&A...594A..86R}), which is typically used for N-body simulations.
To calculate A$_{2}$, the surface density of either the stellar or gas disc
is expressed as a Fourier series through
\begin{equation}
 \Sigma(r,\phi) = \frac{a_{0}(r)}{2}+ \sum_{n=1}^{\infty}\left(a_{m}(r)\cos(m\phi) + b_{m}(r)\sin(m\phi) \right)~,
\end{equation}
where a$_{m}$(r) and b$_{m}$(r) are the radial Fourier coefficients. 
The bar strength is 
\begin{equation}
 A_{2} = \sqrt{a_{2}^{2} + b_{2}^{2}}~.
\end{equation}

We require that both the bar strength and the velocity map of the snapshot match the observations. For the velocity comparison, we first construct a residual map, defined as 

\begin{equation}
 V_{res}^{i} = V_{obs}^{i} - V_{mod}^{i}~,
\end{equation}
where V$_{obs}^{i}$ and V$_{mod}^{i}$ are the observed and modeled velocity at a given pixel {\it i} in the velocity map.
The best-fitting snapshot is the one with the smallest standard deviation $\sigma_{res}$ in the residuals.  This quantity is given by
\begin{equation}
 \sigma_{res} = \sqrt{ \sum_{i=0}^{N}\left(\frac{(\overline{V_{res}^{i}} - V_{res}^{i})^{2}}{N}\right)} ~,
\end{equation}
where N is the number of pixel and $\overline{V_{res}^{i}}$ the mean of the residual velocities.

In addition to these quantities, we also calculate the 'expected' RC for each snapshot and a Fourier decomposition of the radial and tangential gas particle velocities.  These quantities are used to 
compare our results to those obtained using ROTCUR and DiskFit.

The expected circular velocity curve is defined as
\begin{equation}
 \langle V_{expected}^{2}\rangle=\left<r\frac{\partial \Phi}{\partial r}\right> = \left< rF_{r}\right>~,
\label{exp}
\end{equation}
where $F_{r}$ is the radial force from the particles calculated azimuthally 
on an angular grid and $\Phi$ is the gravitational potential. 

The radial and tangential velocity of the gas particles can also be described in terms of Fourier moments.  They are
\begin{equation}
 V_{t}(r,\theta)=A_{0,t}(r) +  \sum_{m=1}^{\infty}A_{m,t}(r)cos[m\theta + \theta_{m,t}(r)]~,
\end{equation}
and
\begin{equation}
 V_{r}(r,\theta)=A_{0,r}(r) +  \sum_{m=1}^{\infty}A_{m,r}(r)cos[m\theta + \theta_{m,r}(r)]~,
\end{equation}
where { \it $A_{m,t}(r)$} and {\it $A_{m,r}(r)$} are the tangential and radial Fourier m$^{th}$ 
velocity moments respectively and $\theta$, $\theta_{m,t}(r)$ and $\theta_{m,r}(r)$ are the angular phases. Describing the velocity fields in terms of Fourier moments allows for a comparison with the non-circular moments found by DiskFit for NGC 1530.

\section{Results and discussions}
\label{sec:Comparison}
In this section, we compare the bar properties and model velocity maps derived from the simulations with the observations. The results for NGC 3621, NGC 1530 and NGC 1300 are presented and discussed in Section \ref{ngc3621}, Section \ref{n1530} and Section \ref{n1300} respectively. 

\subsection{NGC 3621}
\label{ngc3621}
 \begin{figure*}
 \begin{center}$
  \begin{array}{cc}
  \includegraphics[width=170mm]{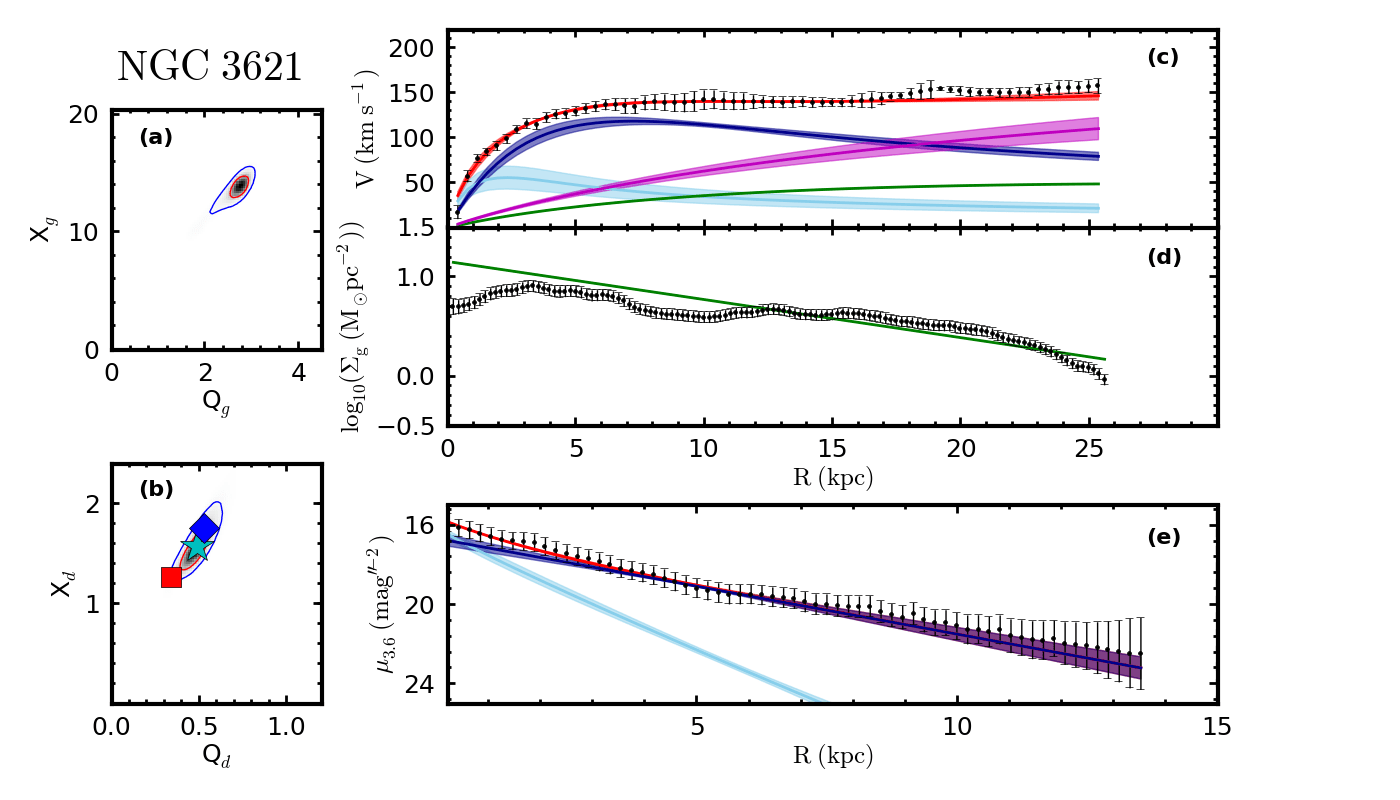}
        \end{array}$
    \end{center}
  \caption{The PDFs of the disc
 stability parameters, $Q$ and $X$, are shown in the left panels. The upper left panel is for the gas disc $(a)$, while the lower left panel $(b)$ is for the stellar disc. The stability parameters for the selected models are plotted on top of panel $(b)$ where Model-A is shown as blue diamond, Model-B as cyan star and Model-C as red square.
The right panels shows the PDFs of the RCs on top $(c)$, the gas surface density on the middle panel $(d)$ and the 3.6-$\mu$ surface brightness profile on the bottom panel $(e)$. The best fit to the data is shown as red lines, the stellar disc is shown as dark blue, the green lines are the gas disc contribution, the magenta line is the halo contribution and the light blue lines are the bulge component. The shaded area shows the 1-$\sigma$ error and the data are shown as black points.}
 \label{fig1}
\end{figure*} 

\begin{figure*}
 \begin{center}$
  \begin{array}{cc}
   \includegraphics[width=90mm]{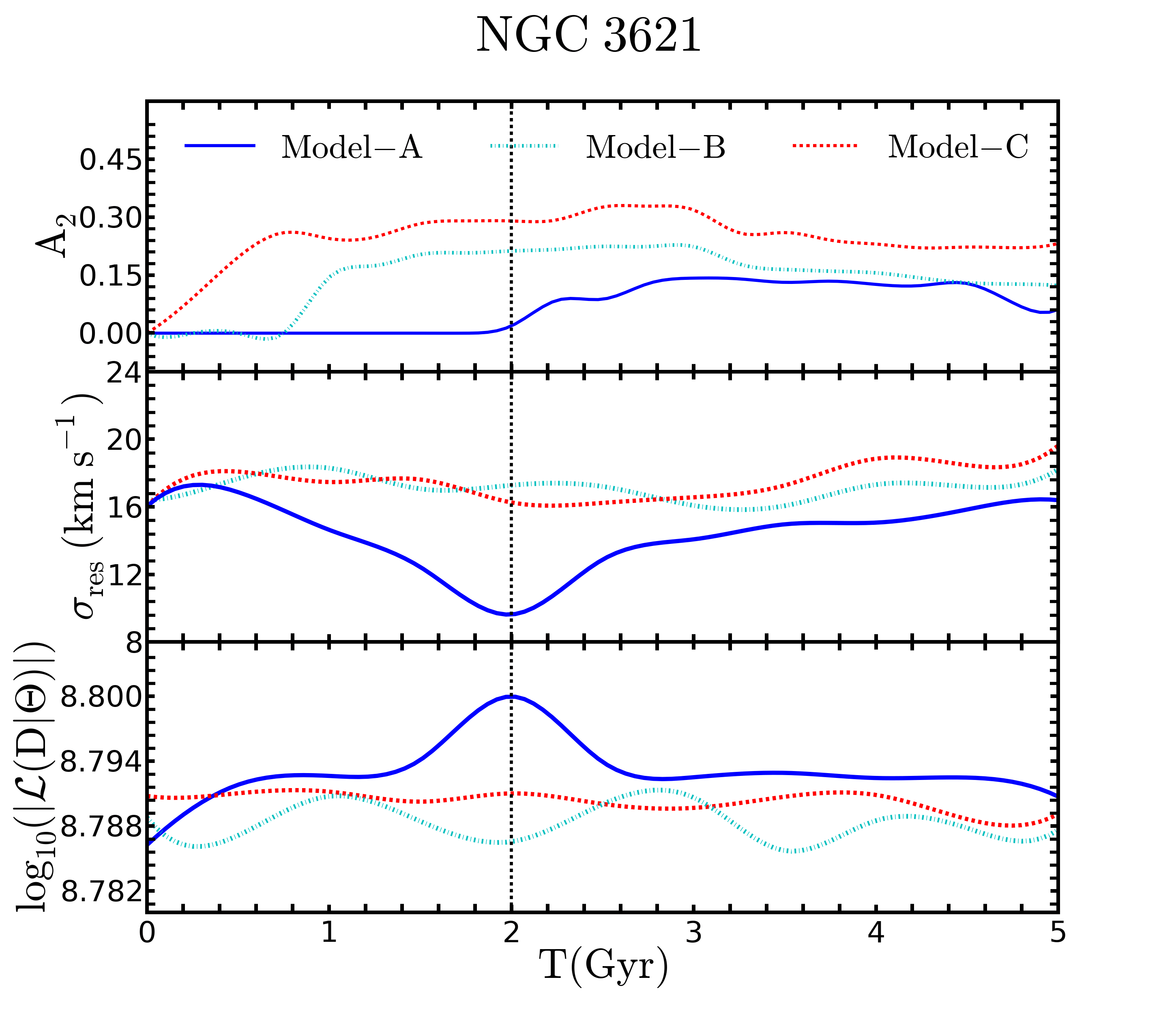}\quad
  \includegraphics[width=95mm]{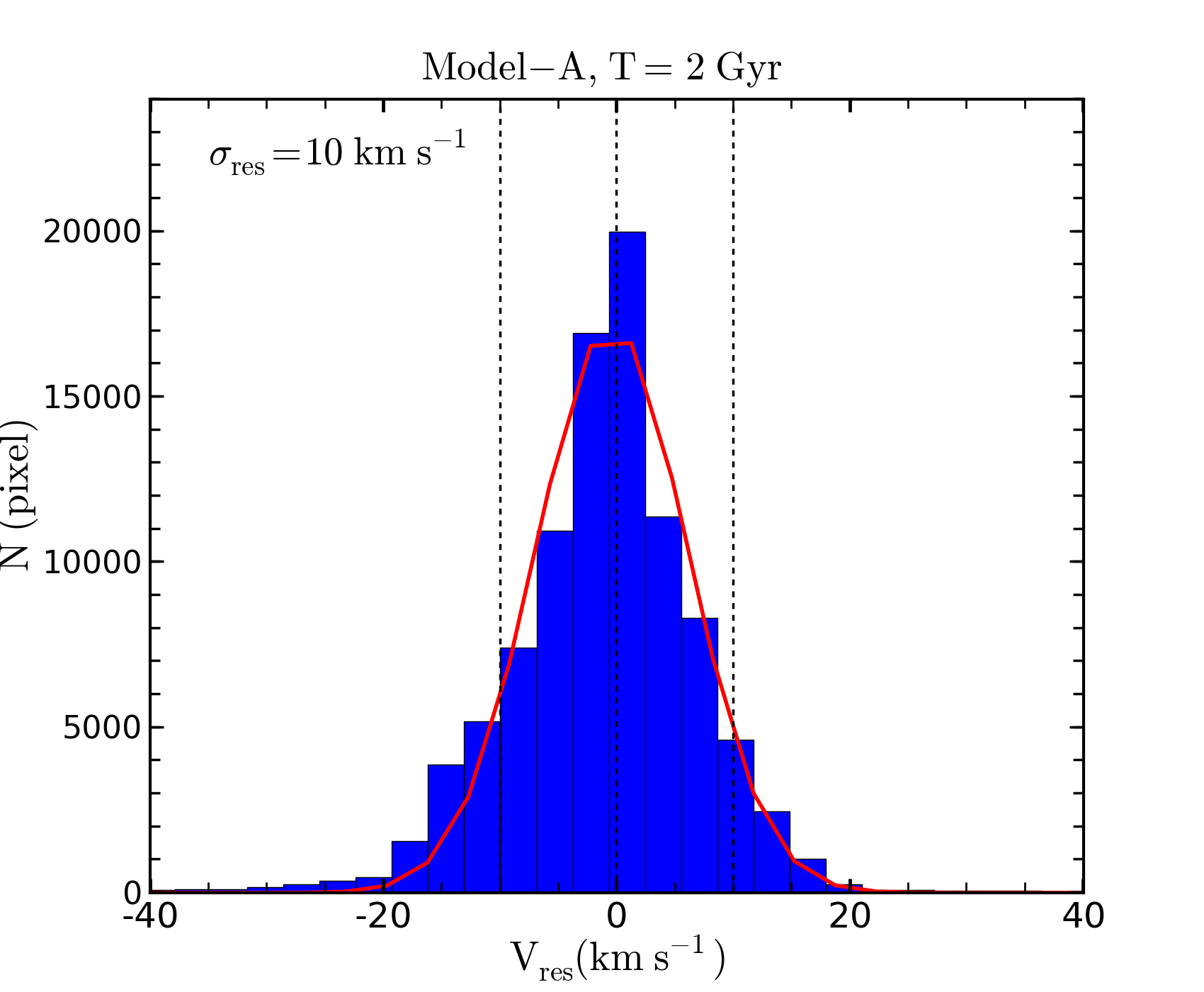}\\
        \end{array}$
    \end{center}
  \caption{Left panels: the variation of the bar strength A$_{2}$ as function of the epoch is shown on the top panel, the standard deviation of the residual $\sigma_{res}$ on the middle and the variation of the likelihood. The vertical dashed line indicates the epoch of the selected snapshot. Right panel: histogram of the residuals for Model-A at T=2.0 Gyr, the red line is the best fit to the data and $\sigma_{res}$ is the standard deviation of the residuals. The vertical dashed lines indicate the mean and standard deviation of the residuals}
 \label{fig2}
\end{figure*}

 \begin{figure*}
 \begin{center}$
  \begin{array}{cc}
  \includegraphics[width=185mm]{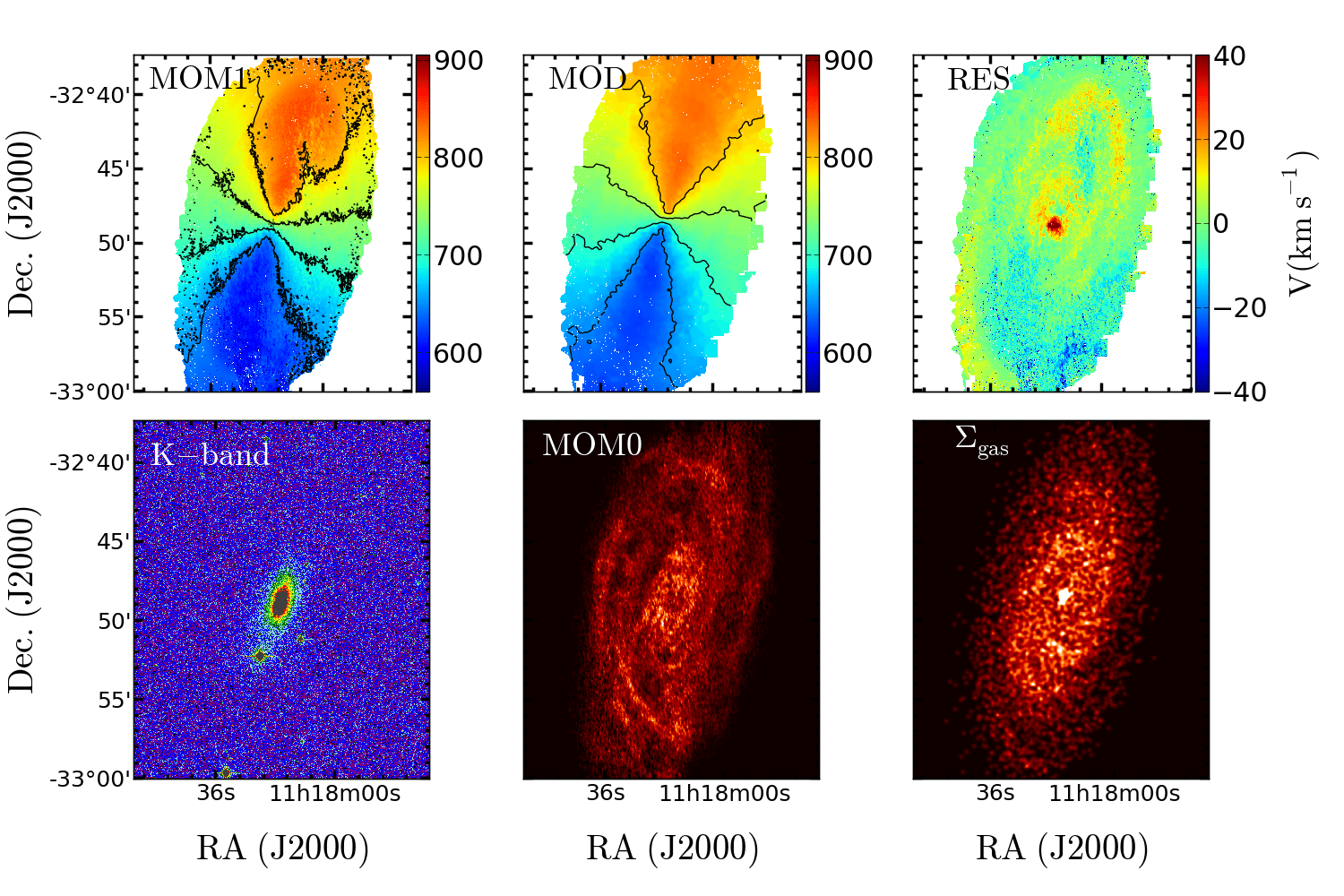}
        \end{array}$
    \end{center}
  \caption{Observed and simulated maps for NGC 3621, Model-A at T=2 Gyr: the moment-1, model velocity field and residual maps are presented on the top panels. The black contours are spaced by 50 km s$^{-1}$. The near-infrared K-band image from \citet{2003AJ....125..525J} (it is important to note that this is not used in any analysis, but to show the corresponding stellar systems to the gas maps that are used for the analysis) is shown with the moment-0 and the simulated gas surface density maps on the bottom panels. The moment1 and moment0 maps are obtained from the THINGS database (\url{http://www.mpia.de/THINGS/Data.html}).}
 \label{fig3}
\end{figure*}

%

 
 \begin{figure}
 \begin{center}$
  \begin{array}{cc}
  \includegraphics[width=80mm]{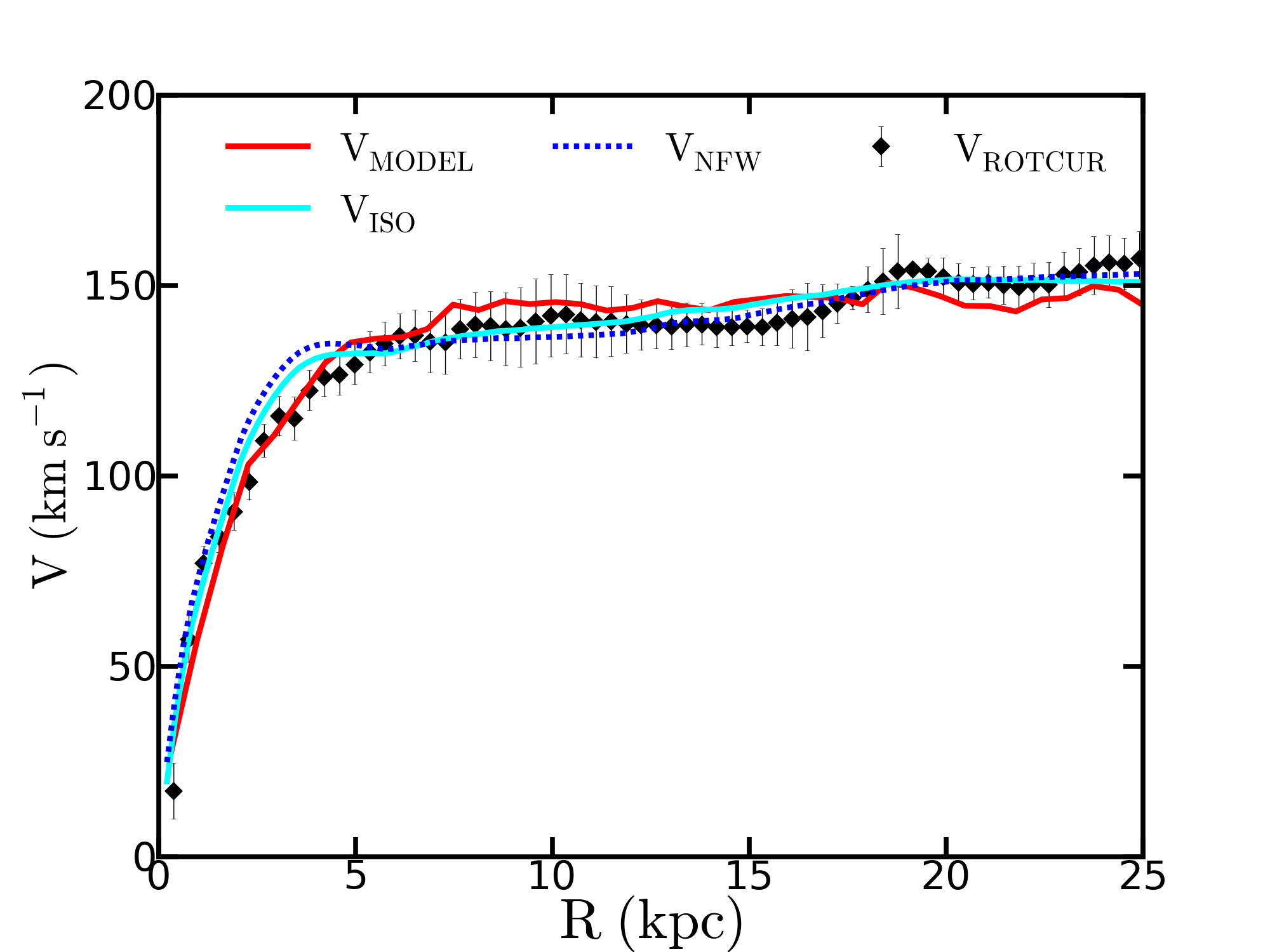}
        \end{array}$
    \end{center}
  \caption{Comparison between the observed RC, the RC calculated from the gravitational potential with the ISO and NFW mass model results from \citep{de-Blok:2008oq} for the NGC3621-A at T = 2.0 Gyr.}
 \label{fig4}
\end{figure}
The Bayesian analysis of the RC, 3.6$\rm \mu m $ surface brightness, and gas surface density of NGC 3621 gives the posterior distribution function (PDF) for the full model parameter space. 
The best fit GalactICs parameters are given in table \ref{tab:3} where the uncertainties are the 1-$\sigma$ error obtained from the PDFs. Using the parameter PDFs we obtained the PDFs of the RC, surface density, and surface brightness profiles shown in Fig. \ref{fig1}, as well as PDFs of the stability parameters. The bulge and disk mass-to-light ratio are also fitted in addition to the GalactICs parameters and the best fit results are given in Table \ref{tab:3}. The bulge and disk mass-to-light ratio are also fitted in addition to the GalactICs parameters and the best fit results are given in Table \ref{tab:3}.
Our best fit result for the disk mass-to-light ratio is consistent with  \citet{de-Blok:2008oq} if it is compared with their mass-to-light ratio obtained using the Kroupa IMF but two times smaller than their mass-to-light ratio measured using the diet Selpeter IMF. While our model has a smaller disk mass, we also have a bulge component.  The sum of the stellar mass is consistent with the \citet{de-Blok:2008oq} results.

Finally, Table \ref{tab:3} also shows the parameters of both the NFW and ISO halo models from de Blok et al. (2008) for comparison.  However, our analysis excludes both cored and cuspy halos, preferring an inner slope of 0.5.  
A better comparison of the halo slopes is to the results of \citet{2011AJ....142..109C}, who used Einasto profiles \citep{1969AN....291...97E,1968PTarO..36..414E,1965TrAlm...5...87E} to model the dark halo of THINGS galaxies.  Like the double power-law profile, the Einasto profile also allows for cored and cuspy halos.  Using a diet-Saltpeter IMF, \citet{2011AJ....142..109C} found an inner slope of 0.53, which is consistent with our PDF.
In addition, our inferred gas mass is slightly larger than the THINGS gas mass (see  \citealt{2008AJ....136.2563W}), 
but this difference is largely due to the different methodologies adopted. The THINGS analysis sums up the mass of all visible gas, 
while we fit an exponential surface density where M$_{g}$ is the total integrated mass of that exponential.  
Nonetheless, Fig. \ref{fig1} shows that our model provides good fits to the data and is comparable with both the NFW and 
ISO models of \citet{de-Blok:2008oq} (see Fig \ref{fig4} ).

While NGC 3621 is classified as an unbarred galaxy, it is unclear which models will form bars a priori.  Therefore we selected three models with different disc stability parameters. This selection allows for a brief exploration of the parameter space. These models (A,B, and C) are listed in Table \ref{tab:3} and plotted in Fig. \ref{fig1}.  Model A has the smallest disc/halo ratio, which should give it the most stability against bar formation. Model B correspond to the peak of the parameters PDFs and Model-C is the least stable. 
After evolving the model for 5 Gyr, we extract the stellar surface density and gaseous velocity maps for every snapshot. 
At each time-step we calculated both the bar strength and the standard deviation of the difference between the simulated and 
observed velocity maps, which are shown in the left panels of Fig. \ref{fig2}. The standard deviation is obtained first by 
re-centering and re-aligning the model and observed velocity fields using the GIPSY task TRANSFORM. Then a pixel-by-pixel 
comparison is performed using the GIPSY utility task SUB. The variation of the likelihood is also shown on the bottom 
left panel of Fig. \ref{fig2}. The likelihood is given by equation \ref{equa7}, except O$_{i}$ is the observed MOMENT1 map, M$_{D,i}$ 
the model velocity fields and $\sigma_{D,i}$ the observed MOMENT2 map, which is adopted as the velocity field uncertainties. 
While we do not define our best-fitting model by the likelihood, it does provide confirmation that our residual method selects 
the best fitting snapshot from the set of simulations. The right-hand panel of Fig. \ref{fig2} shows the histogram of the 
residuals from the best-fitting snapshot.  The left-hand panels shows that Model A has the weakest bar at all time steps, and 
at T=2 Gyr, it has the lowest standard deviation of all models.  At that same time step, the likelihood is maximized.

Fig. \ref{fig3} shows the comparison of the selected snapshot with the observed maps. NGC 3621 exhibits a strong warp in the outermost regions.  Unfortunately our simulations are unable to generate warps at this time, so we exclude that region from our analysis and have focused on matching the central, axisymmetric portions of the galaxy
Nonetheless, our method succeeds at our goal of modeling the central portions of the disk, obtaining residuals in the innermost regions of $\pm$ 10 km s$^{-1}$. While the velocity maps show a good fit for the model, it is clear that there are differences between the Mom0 image and our simulated gas density.  This is due to our simulations using an exponential gas disk that cannot reproduce the central holes observed in the H{\sc i} disk.

It is interesting to note that Model-A is a high stability model, not the model found at the peak of the parameter PDFs. This result emphasizes the importance of running numerical simulations as the Bayesian PDFs include model that rapidly form moderate bars. While these initial conditions are acceptable for the axisymmetric Bayesian analysis, they must be rejected based on the simulations.



Fig. \ref{fig4} shows a comparison of the ROTCUR RC with the THINGS model RCs and the 
expected circular velocity curve calculated from our best fitting snapshot. These RCs are all consistent with the 
ROTCUR RC. Interestingly, the simulated RC seems to match the ROTCUR inner 5 kpc slightly better than the THINGS RCs, but 
it is also slightly below the ROTCUR results in the outermost regions.  However, our goal is not to match the 
RC, but to match the observed velocity map.  Nonetheless,
our analysis of NGC 3621 is at least as accurate as ROTCUR and other 
standard methods at recovering the RC for an axisymmetric model.  As an added bonus, our RC is built from the
underlying mass model, so we do not need to perform an additional inference to recover the dark halo profile.
 
 

\subsection{NGC 1530}
\label{n1530}

\begin{figure*}
 \begin{center}$
  \begin{array}{cc}
  \includegraphics[width=170mm]{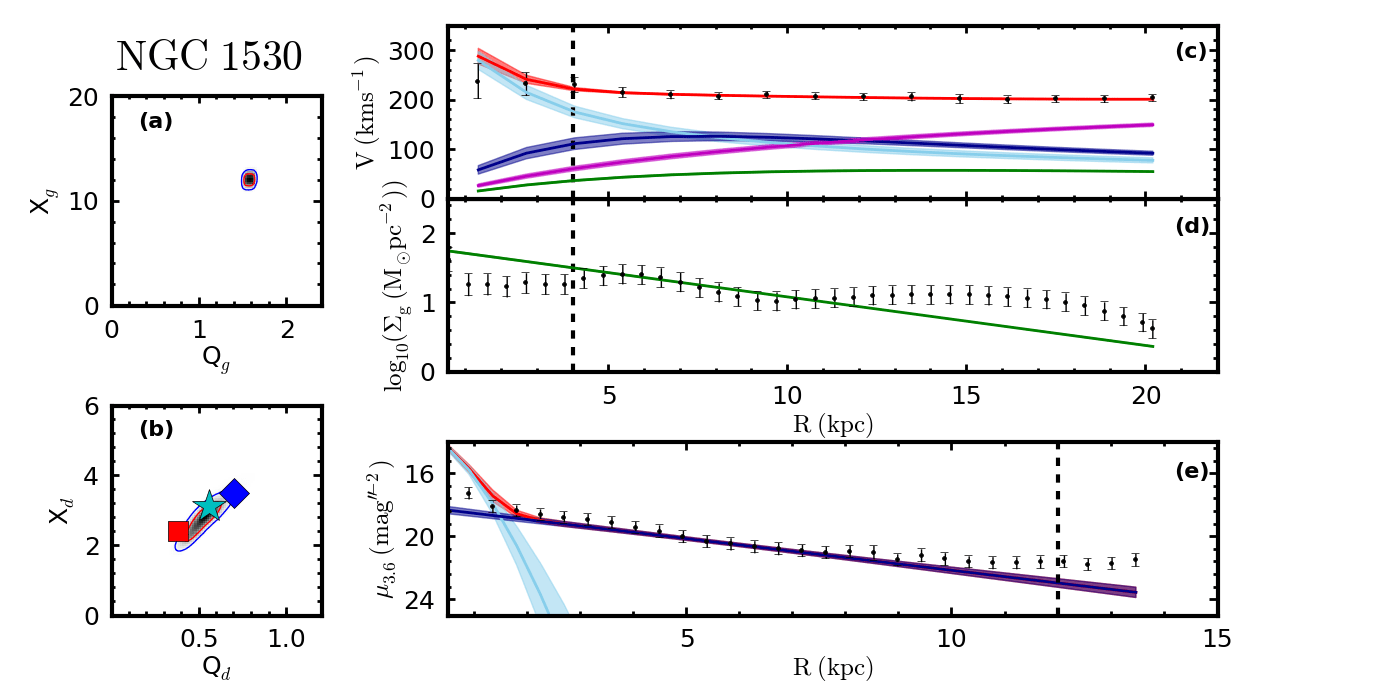}
        \end{array}$
    \end{center}
  \caption{Same as in Fig. \ref{fig1} but for NGC 1530. The vertical dashed lines indicate regions that are excluded from our analysis due to the presence of the bar (RC and gas SD) or possible contamination (surface brightness).}
 \label{fig5}
 \end{figure*} 
 
 \begin{figure*}
 \begin{center}$
  \begin{array}{cc}
   \includegraphics[width=90mm]{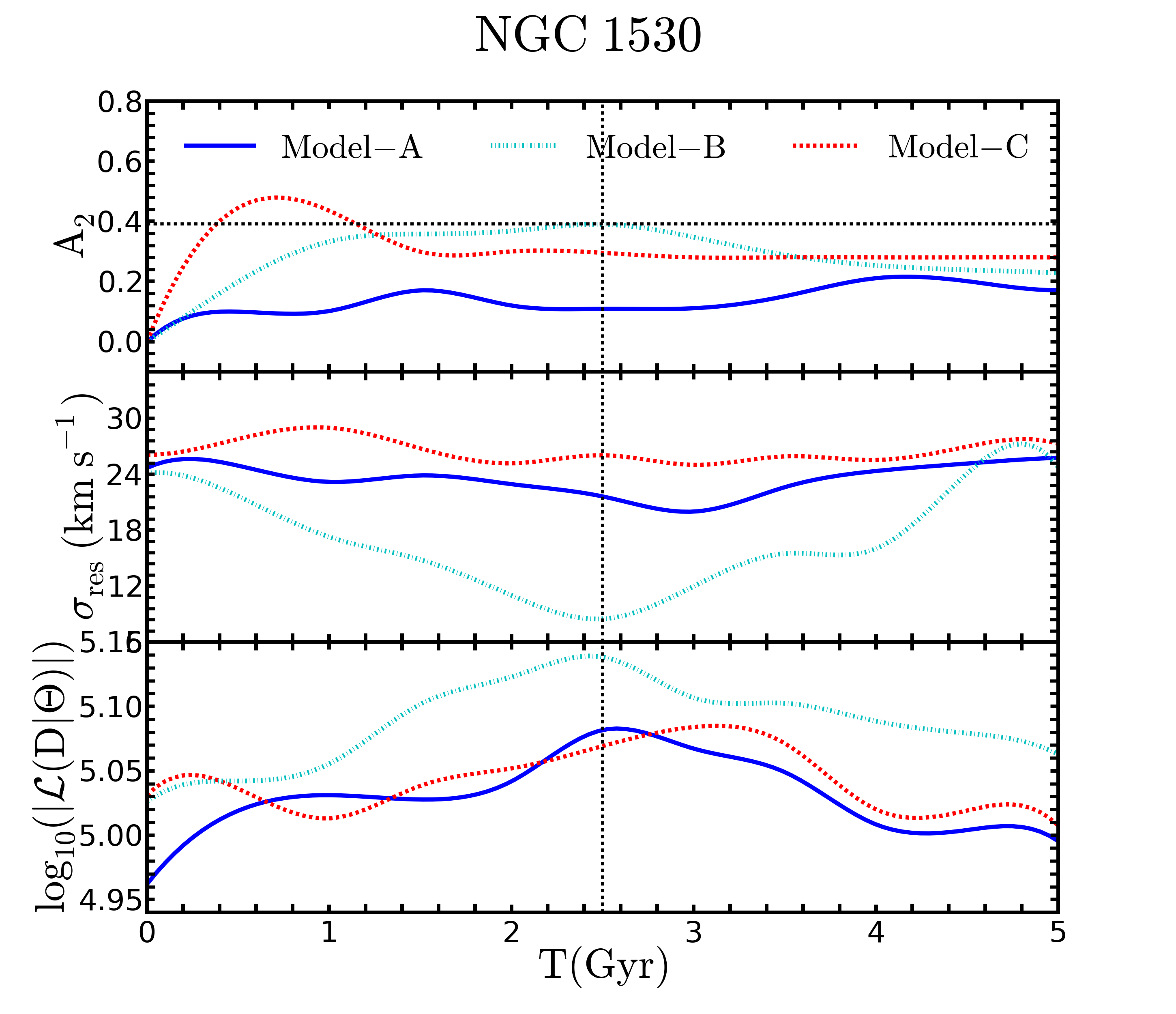}\quad
    \includegraphics[width=95mm]{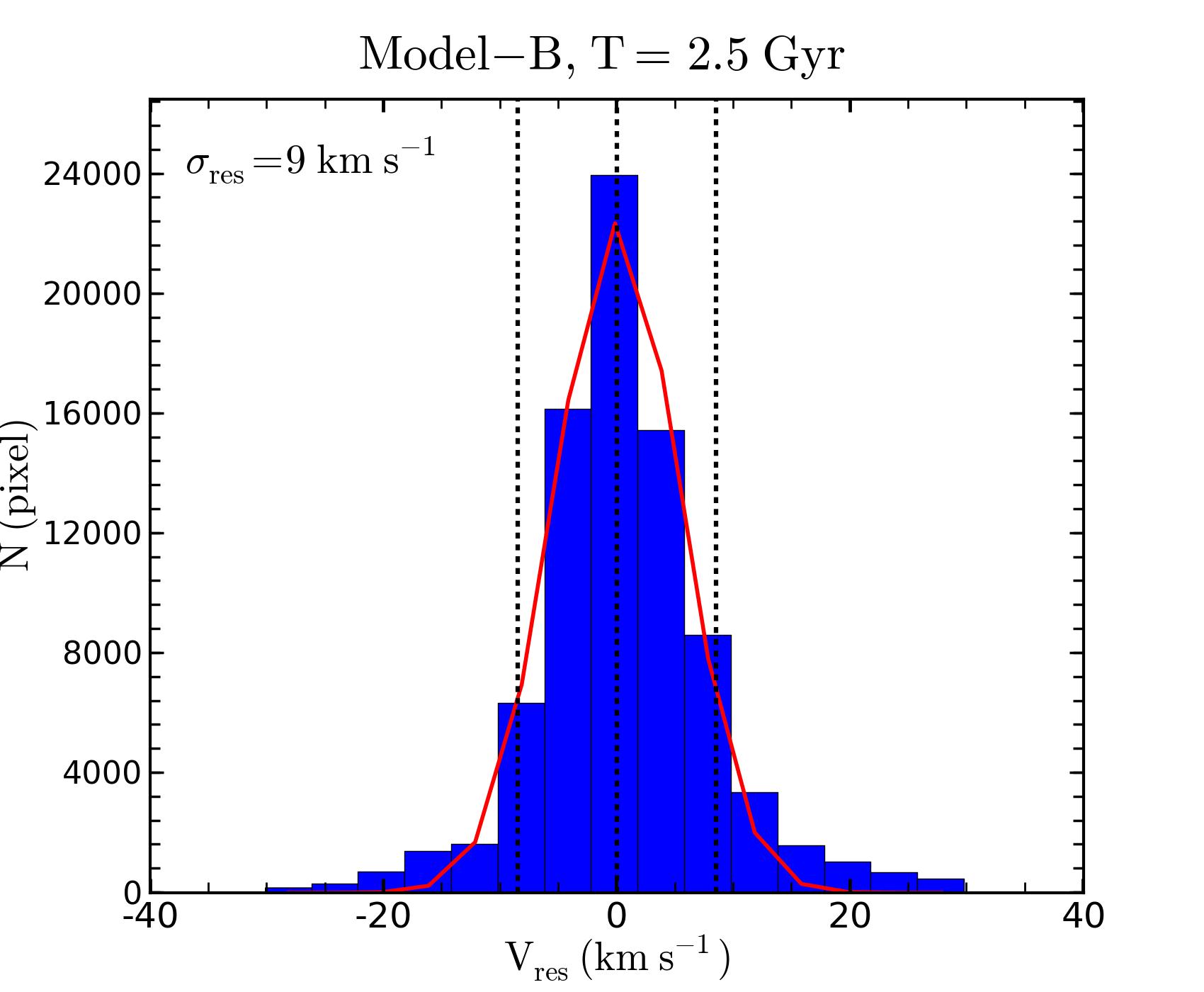}\\
        \end{array}$
    \end{center}
  \caption{Same as Fig \ref{fig2} for NGC 1530. The dashed horizontal line is the A$_{2}$ from \citet{1998AJ....116.2136A}. The vertical dashed line indicates the location of the selected snapshot.}
 \label{fig6}
\end{figure*}

 \begin{figure*}
 \begin{center}$
  \begin{array}{cc}
  \includegraphics[width=190mm]{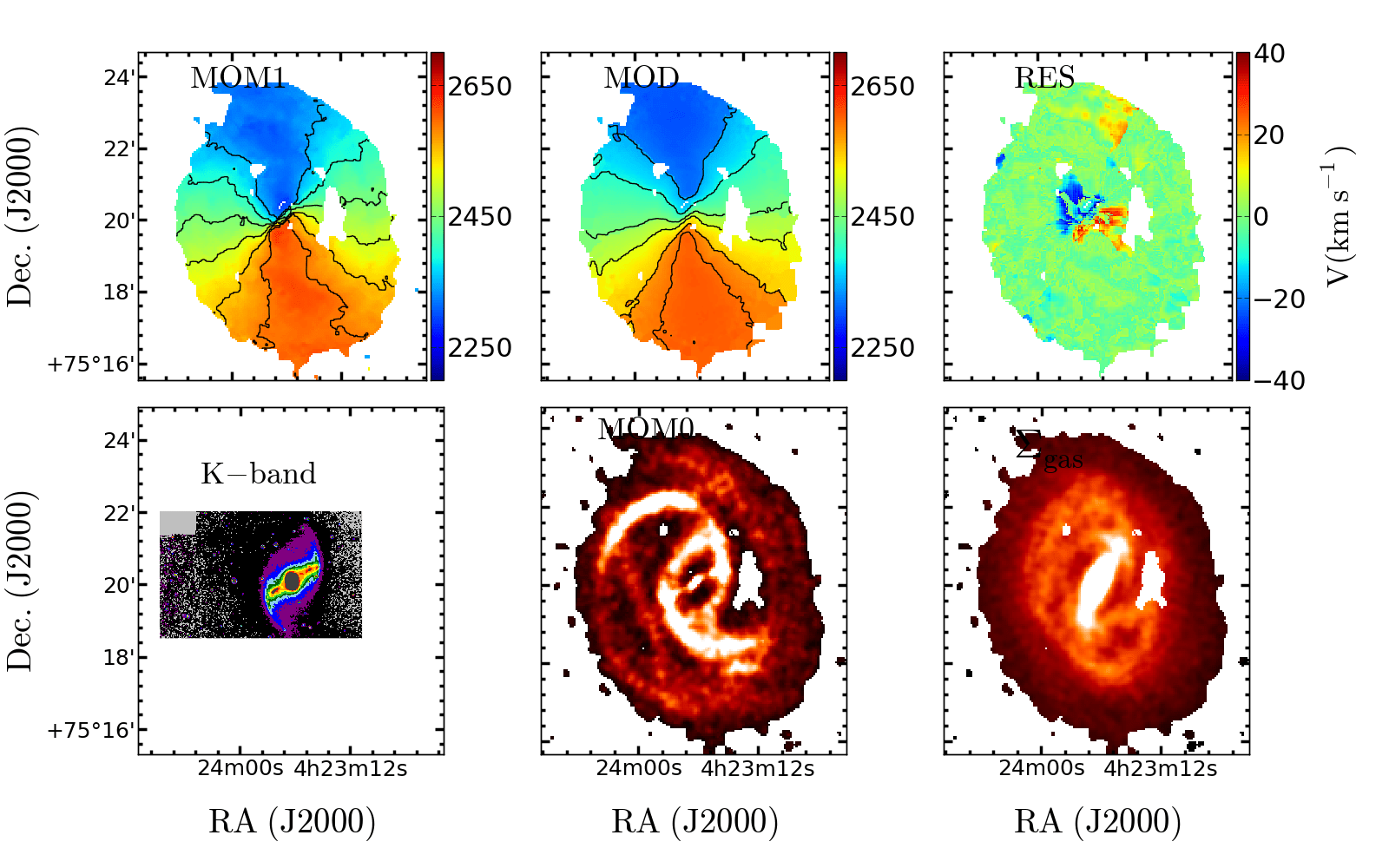}
        \end{array}$
    \end{center}
  \caption{ Simulated and Observed maps for NGC 1530, Model-B. The panels are the same  as in Fig. \ref{fig3}. The black contours are spaced by 50 km s$^{-1}$. The near-infrared K-band image is taken from \citet{1995ApJ...449..576R}.}
 \label{fig7}
\end{figure*}

%
%

\begin{figure*}
 \begin{center}$
  \begin{array}{cc}
  \includegraphics[width=65mm]{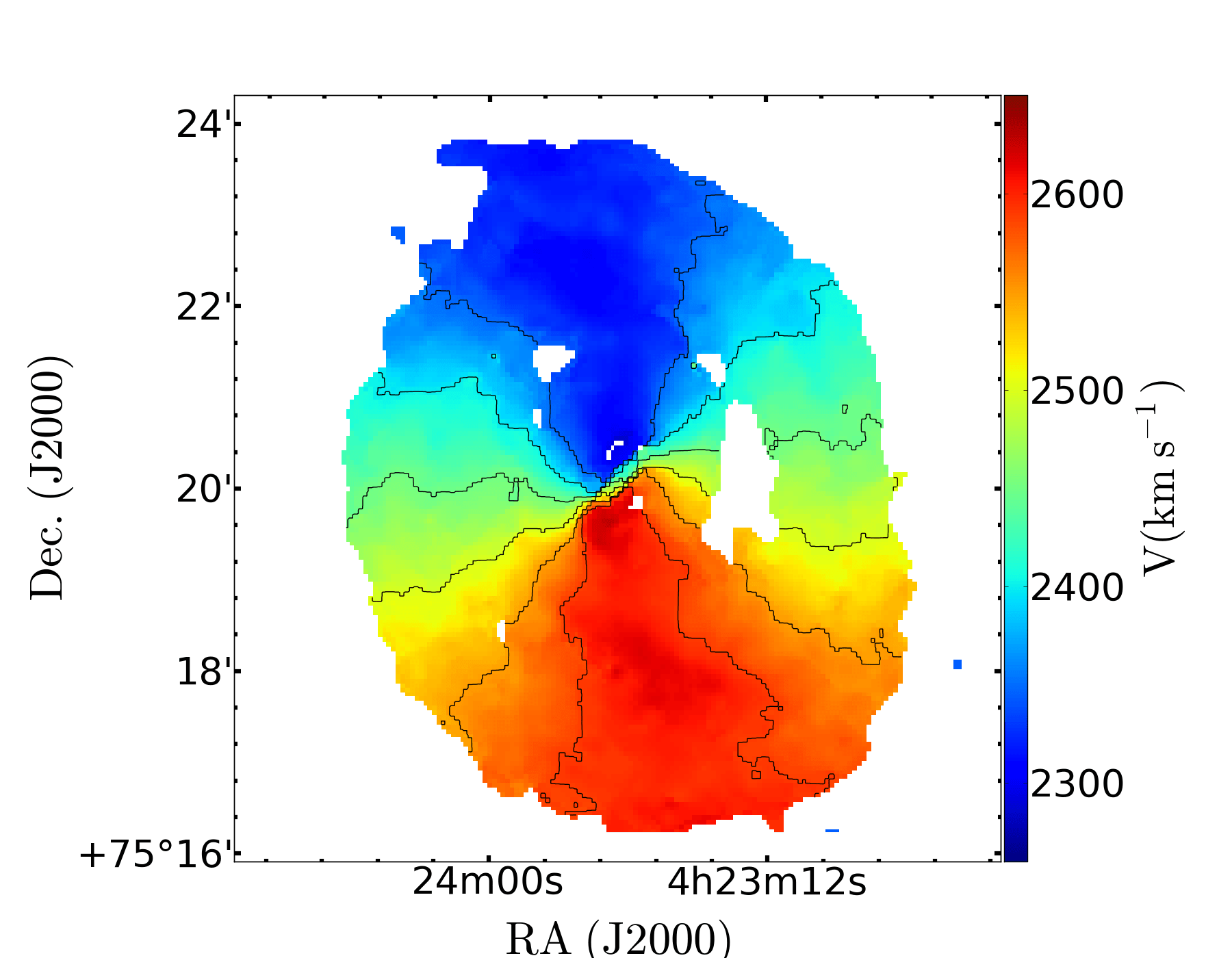}
  \includegraphics[width=65mm]{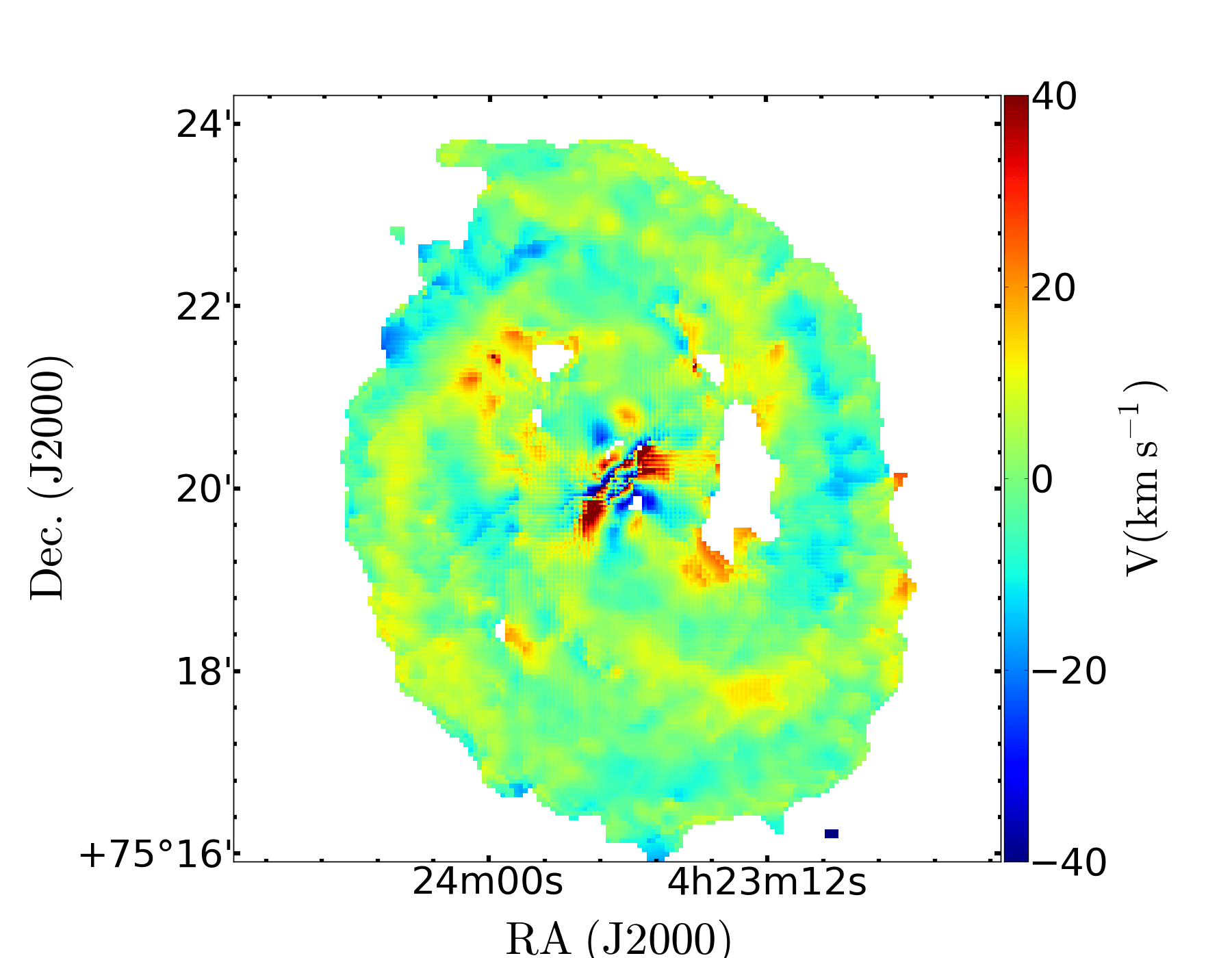}\\
  \includegraphics[width=65mm]{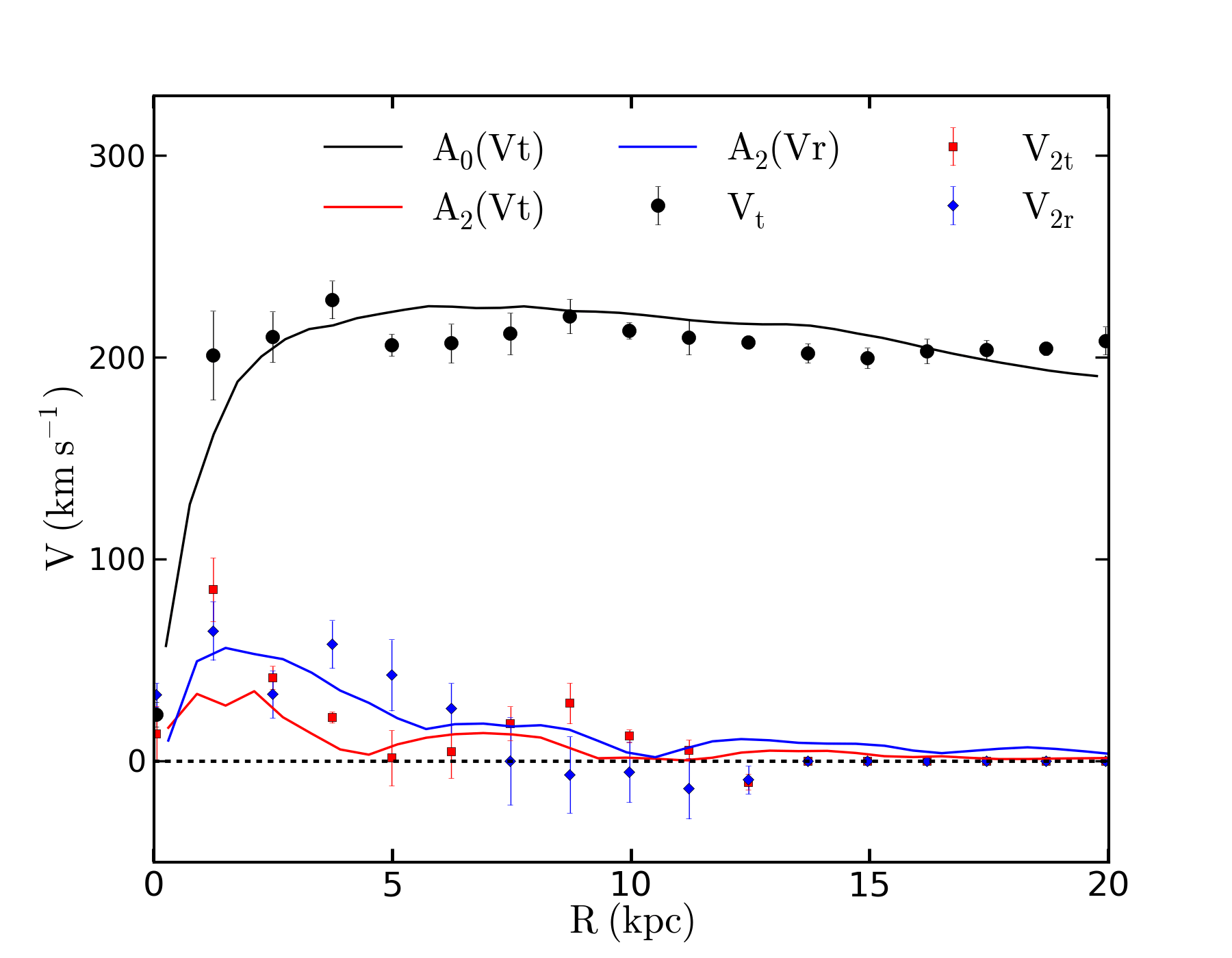}
  \includegraphics[width=65mm]{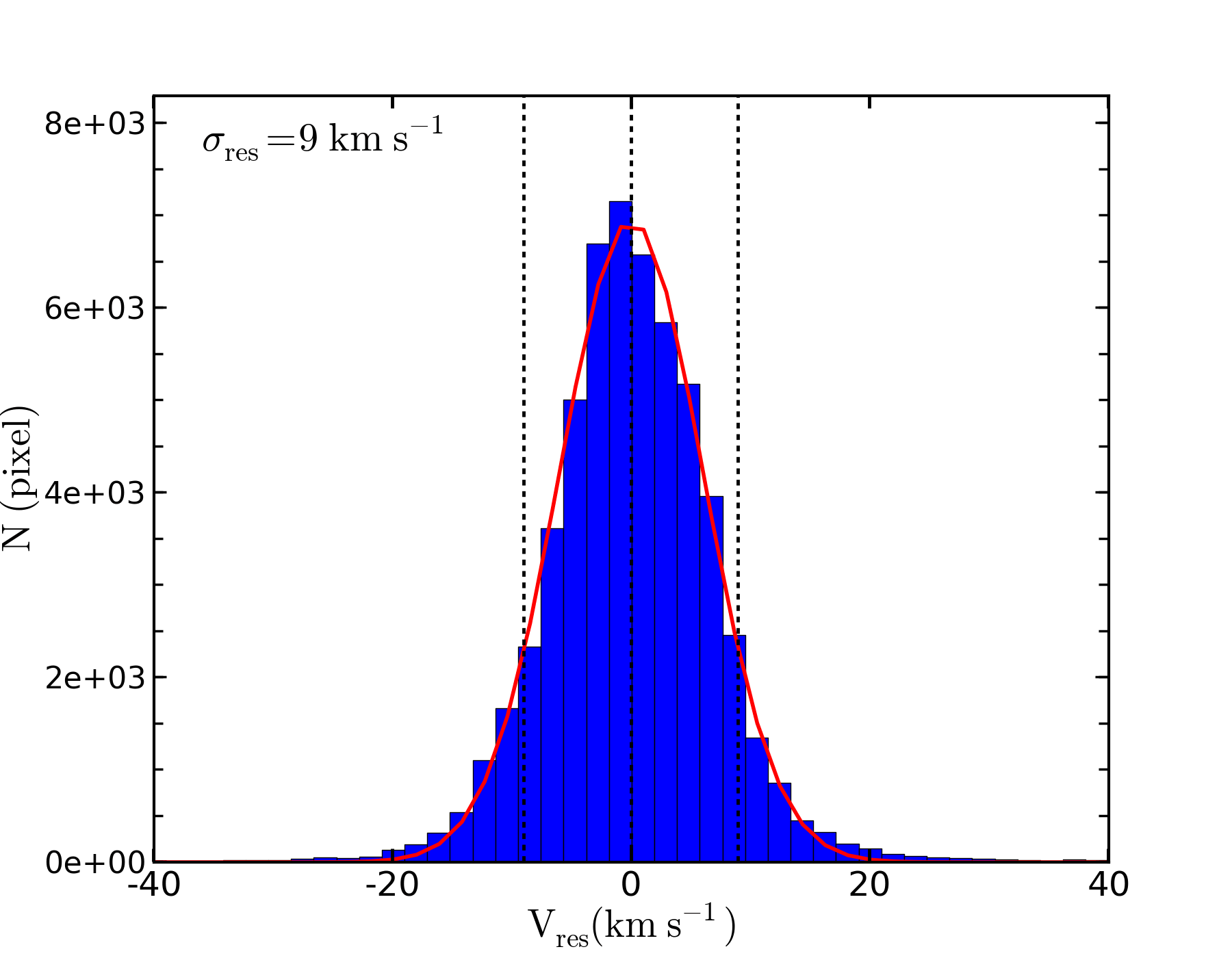}
        \end{array}$
    \end{center}
  \caption{DiskFit results for NGC1530. The moment1 map is on the top left panel and the residual map is displayed on the top right panel. A histogram of the residuals is presented on the bottom right panel and a comparison between the DiskFit velocities $\rm V_{t}$, $\rm V_{2t}$ and $\rm V_{2r}$ with the amplitude of the m=0 and m=2 Fourier mode $A_{0}$ and $A_{2}$ on the left panel. The black contours are spaced by 50 km s$^{-1}$.} 
 \label{fig8}
\end{figure*}


NGC 1530 provides an excellent test case for comparing our method with DiskFit as it has 
a strong bar at an intermediate angle.  The best fit results from the initial Bayesian analysis 
are presented in Table \ref{tab:4} while Fig. \ref{fig5} shows the model stability and the comparison to the observations.  In this case we avoid the barred regions of the galaxy for the Bayesian analysis in the RC and surface density fits while we avoid the outer regions of the 3.6 $\mu$m surface profile due to possible contamination.

As with NGC 3621, we simulated three models that represent the parameter space. 
These models correspond to low and high disc stabilities and the peak of the parameter PDFs. 
The initial conditions for the simulation are listed in Table \ref{tab:4}. We analyzed each snapshot, 
comparing both the bar strength, the standard deviation of the velocity residuals and the likelihood for each model, 
which is shown in the left panels of Fig. \ref{fig6}. 
We then selected the model and snapshot that not only reproduces the bar strength but also has the lowest standard deviation of the residual.  One key issue with NGC 1530 that is question of the bar orientation.  The K band image has the optical bar oriented at -25 degree.  However, the Mom0 gas map shows a double-bar structure with the inner bar oriented along the stellar bar, but the outer bar is at -35 degree.  We have explored a number of model orientations and find the best results when we align the model with the outer gas bar. 

The evolution of the bar strength and residuals shown in Fig. \ref{fig6} demonstrates a number of 
interesting points. Model C shows rapid bar growth until it buckles at T $\sim$ 0.8 Gyr and decreases to a constant value. 
Interestingly the minima of Model C does not occur when it has the correct bar strength. On the other hand, Model 
A never grows a bar that is as strong as the observed bar of NGC 1530.  Only Model B has a minima when it matches the bar strength.  
This minima is also the global minima, meaning that the snapshot of Model B at T=2 .5 Gyr is our 
selected model.  As shown in the right-hand panel of Fig. \ref{fig6}, this
snapshot has standard deviation $\sigma_{res}=9~\kms$.


Fig. \ref{fig7} shows the comparison of this snapshot to observations of the system. The largest residuals 
occur near the center of the velocity field. 
However, the MOM0 map shows a greater degree of structure than is seen in the K-band image. This additional structure may explain some of the residual structure seen in Fig. \ref{fig7}.  To 
fully explore this issue, additional simulations will be required.

  The residual map obtain from DiskFit analysis of the same velocity map shown in Fig.  \ref{fig8} is comparable to our method, even though  DiskFit 
is explicitly designed to match velocity maps of galaxies like NGC 1530. Furthermore, as shown in the lower left panel of Fig.  \ref{fig8}, our inferred RC and non-axisymmetric motions agree with those obtained by 
the DiskFit analysis.  While DiskFit is designed to reproduce the velocity maps of barred galaxies, our results remain  
comparable.  Moreover, we can apply our method to galaxies like NGC 1300 and, as mentioned earlier, our method gives a full mass model.



\subsection{NGC 1300}
\label{n1300}

\begin{figure*}
 \begin{center}$
  \begin{array}{cc}
  \includegraphics[width=160mm]{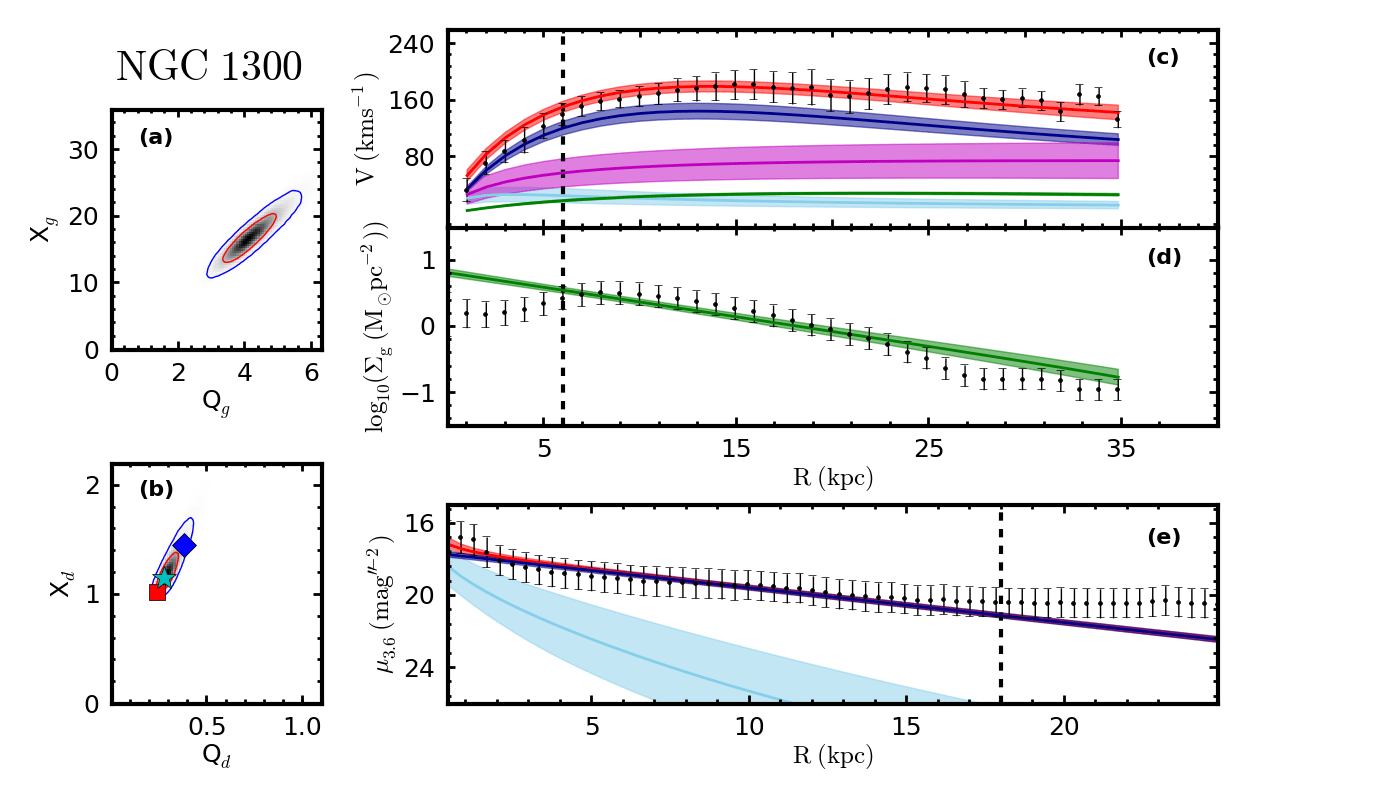}
        \end{array}$
    \end{center}
  \caption{Same as in Fig. \ref{fig1} for NGC 1300, the vertical dashed delineate the part of data used in the fit.}
 \label{fig9}
\end{figure*}  

%

%

\begin{figure*}
 \begin{center}$
  \begin{array}{cc}
   \includegraphics[width=90mm]{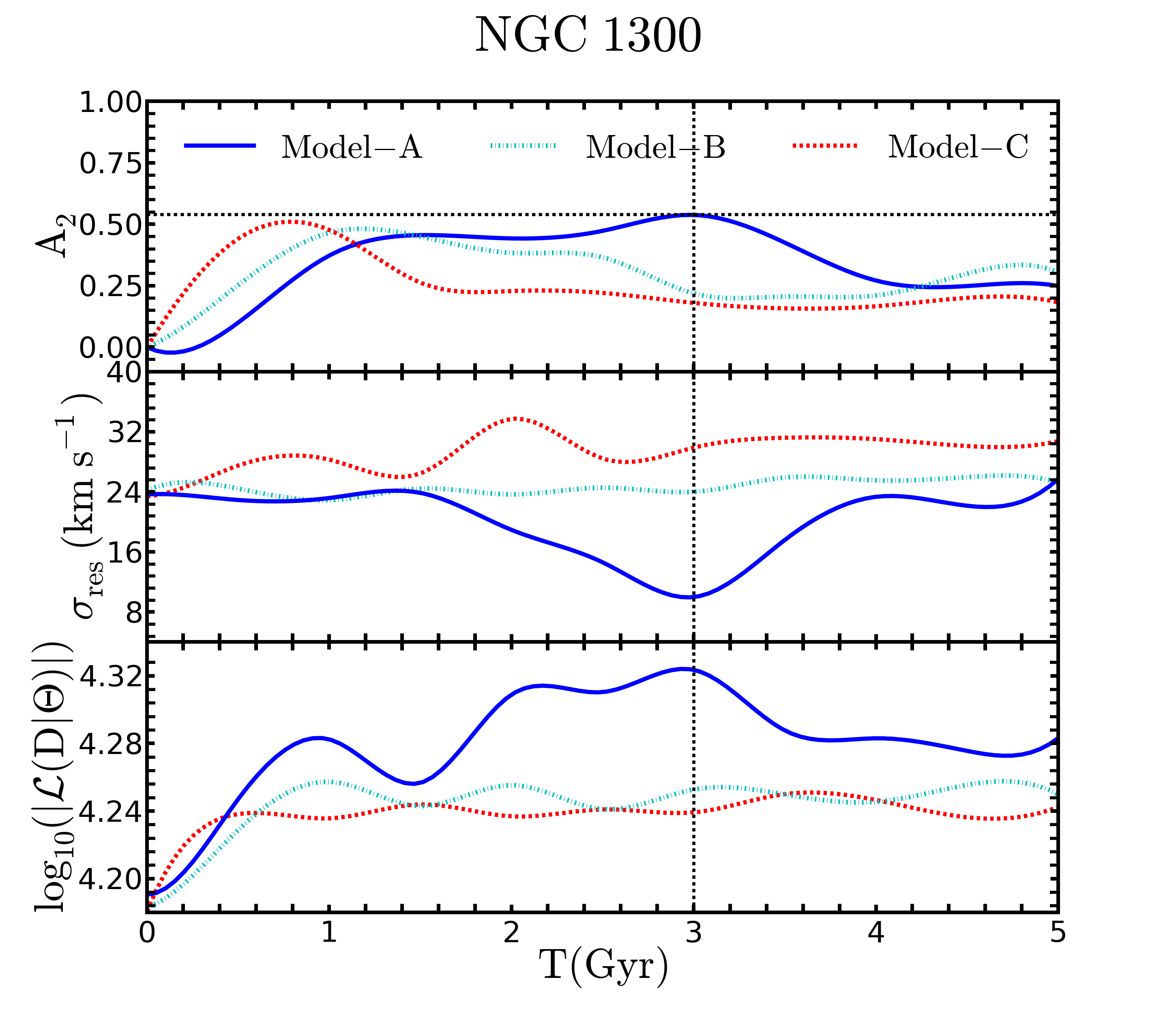} \quad
  \includegraphics[width=100mm]{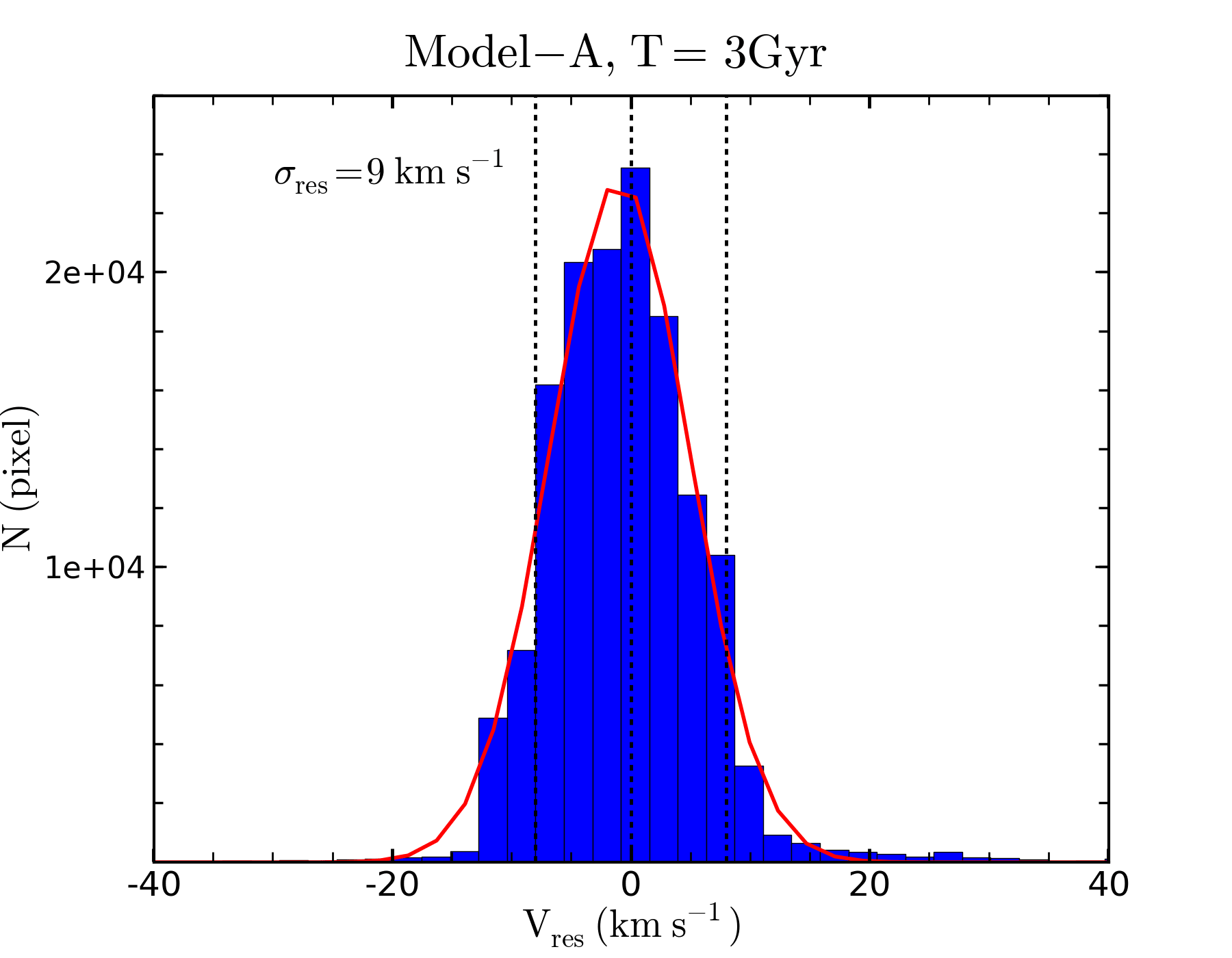}\\
        \end{array}$
    \end{center}
  \caption{Same as Fig \ref{fig2} for NGC 1300. The dashed horizontal line is the A$_{2}$ from \citet{2016A&A...587A.160D}. The vertical dashed line indicates the location of the selected snapshot.}
 \label{fig10}
\end{figure*}

\begin{figure*}
 \begin{center}$
  \begin{array}{cc}
  \includegraphics[width=185mm]{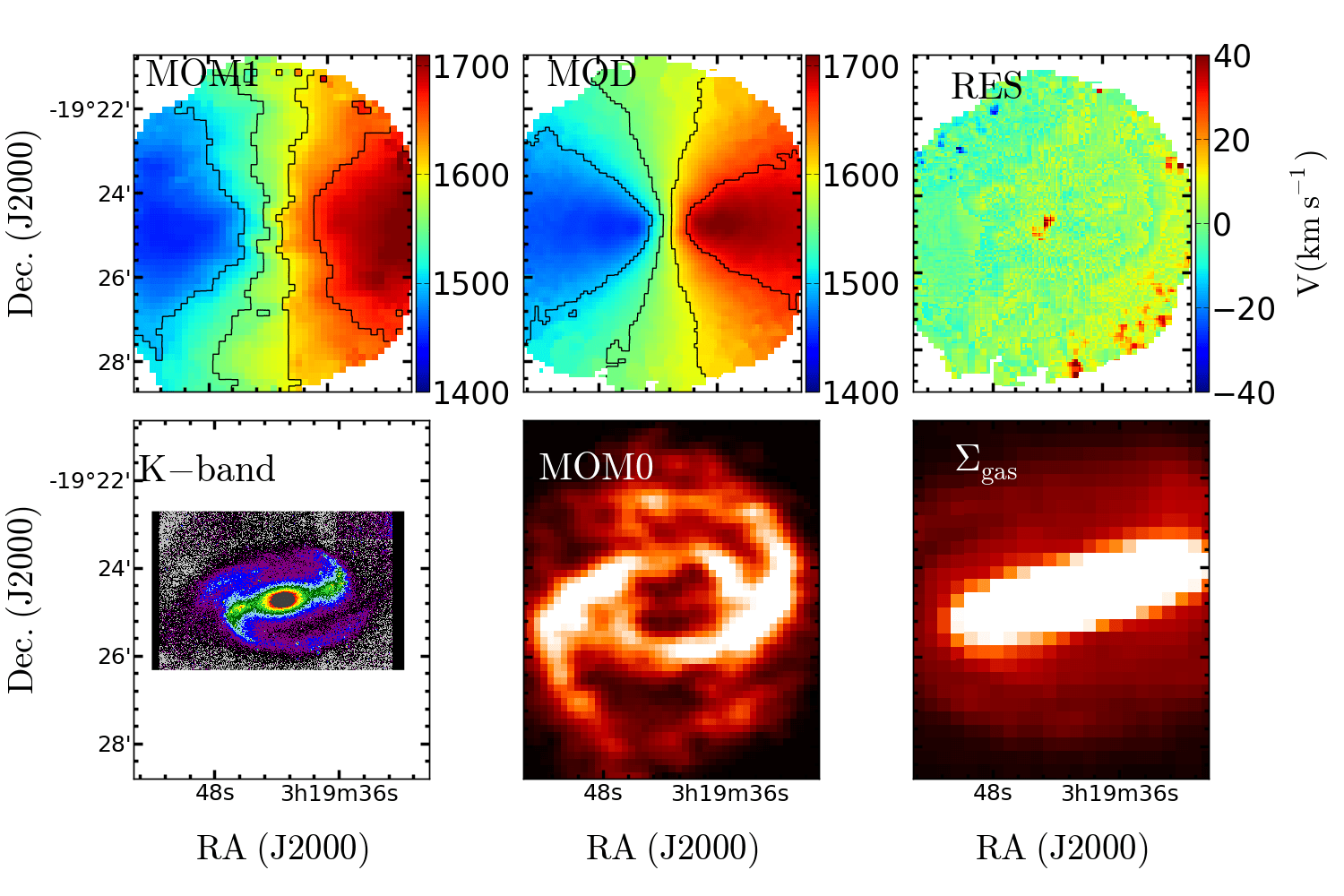}
        \end{array}$
    \end{center}
  \caption{Simulated and Observed maps for NGC 1300, Model-A. The panels are the same  as in Fig. \ref{fig3}. The black contours are spaced by 50 km s$^{-1}$.  The near-infrared K-band image is taken from \citet{2003AJ....125..525J}.}
 \label{fig11}
\end{figure*}


\begin{figure*}
 \begin{center}$
  \begin{array}{cc}
   \includegraphics[width=140mm]{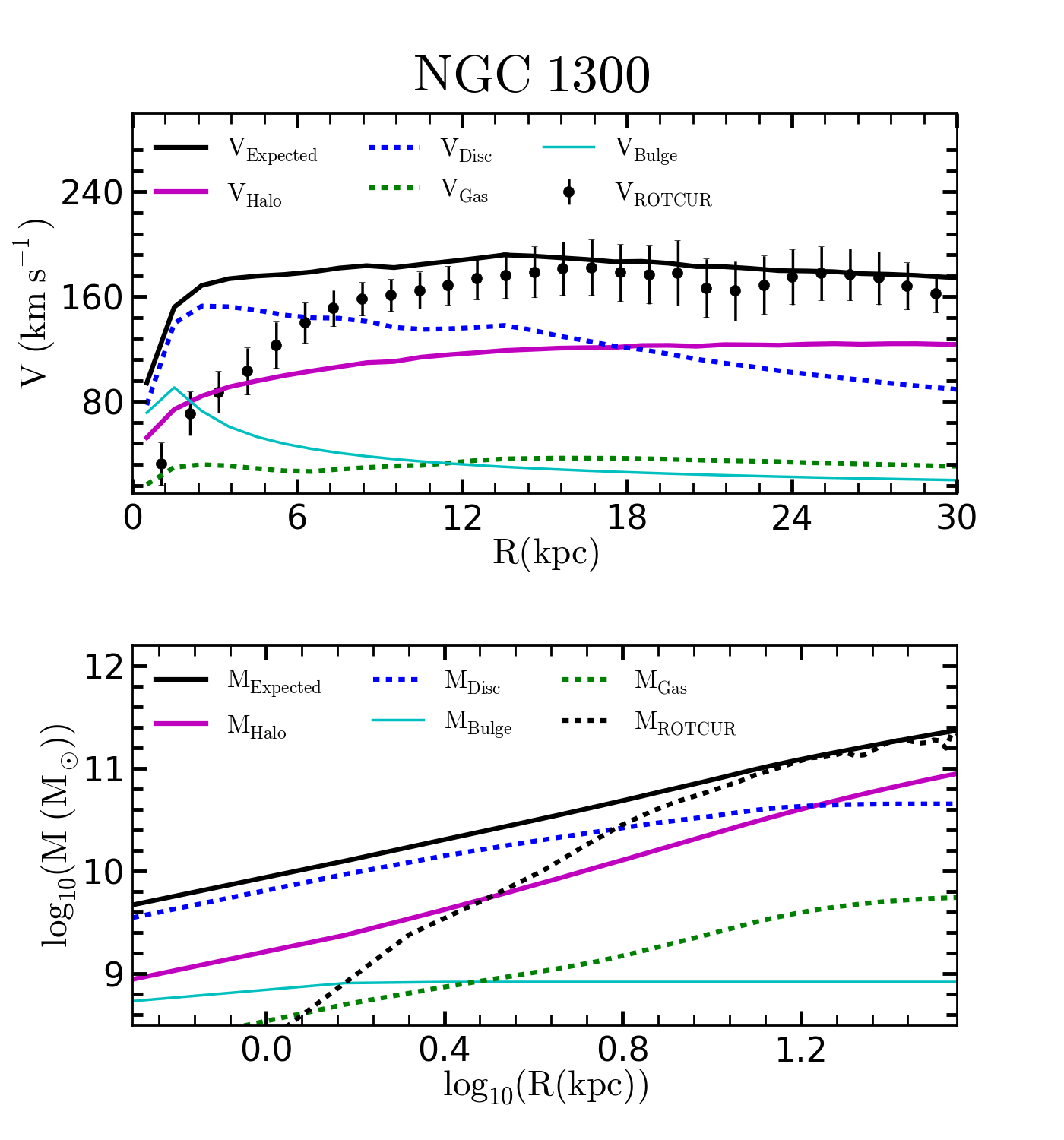}\\
        \end{array}$
    \end{center}
  \caption{The RC measured with ROTCUR is compared with the expected velocities from the gravitational force. The mass profile inferred from the ROTCUR RC and the expected mass from the snapshot are presented on the bottom panel where the black points are the ROTCUR RC, the expected velocities and mass are shown as black lines, the stellar disc is shown as dashed blue, the green dashed line is the gas disc contribution, the magenta line is the halo contribution, the cyan line the bulge component and the dashed black line on the bottom panel is the mass profile inferred from the ROTCUR RC.}
 \label{fig12}
\end{figure*}

NGC 1300 is the perfect galaxy for this method of tailored numerical simulations.  It has a strong bar aligned with the major axis, which means that ROTCUR will underestimate the RC of this galaxy (see \citealt{2016A&A...594A..86R}) and, due to degeneracies in the fitting formula, DiskFit will fail \citep{2010MNRAS.404.1733S}.  As with the other galaxies, we begin with a Bayesian analysis of the RC, 3.6 micron surface brightness, and surface density.  To avoid complications due to the bar, we exclude data from the RC and the gas surface density from the inner $6~\textrm{kpc}$.  The best fit parameters are given in Table \ref{tab:5}.  The stability PDFs and observation comparisons are shown in Fig. \ref{fig9}.  This analysis suggests that NGC 1300 has a halo that is more cuspy than cored.

Following our general methodology, the input parameters for the three models, namely A, B and C are are selected based on their disc stability and listed in Table \ref{tab:5}.  We examine the snapshots from our simulations that have the same bar strength as reported in \citet{2016A&A...587A.160D} and the minimum residual velocity standard deviation. Fig. \ref{fig10} shows that each of the simulations reach the desired bar strength over different time scales. However, when Model A reaches the target bar strength at T=3.5 Gyr it also has the smallest residual with  $\sigma_{res}=9~\kms$. This is also the best representation of NGC 1300 according to the likelihood plot shown on the bottom panel of Fig. \ref{fig10}.
The comparison of the best-fitting snapshot to the actual observations is shown in Fig. \ref{fig11}. It is  clear that our methodology 
can be applied to bars that are parallel/perpendicular to the major axis of the galaxy.  And, for the first time, we have produced a 
mass model for NGC 1300. To illustrate the strength of our method, we compared the RC derived by ROTCUR to the expected circular velocity curve from the simulation.  As shown in the upper panel of Fig. \ref{fig12}, there is a significant difference between these two RCs, 
with the ROTCUR curve underestimating the velocity in the inner 10 kpc.  The lower panel of Fig. \ref{fig12} shows the 
mass profile inferred from the ROTCUR RC compared to the true mass profile of the galaxy.  There is a significant difference 
between these two mass profiles, which will lead to radically different inferences of the DM content of NGC 1300. This emphasizes 
the importance of correcting RC of barred galaxies for the non-circular motions induced by a bar before using them for mass model purposes.

\section{ Summary and Conclusions}
\label{sec:summary}

We have presented a two-step method to study the mass distribution of strongly barred galaxies using numerical simulations.  An initial Bayesian analysis is performed on the azimuthally averaged RC, 3.6 $\rm \mu m$ surface brightness, and gas surface density.  The PDFs from the Bayesian analysis are combined with an examination of the resulting disc stability to determine the initial conditions.  This method is suitable for barred galaxies with a bar that is parallel to one of the symmetry axes or when the RC is not reliable to derive the mass distribution.

The N-body systems are initialized using a new version of the GalactICS code and evolved for 5 Gyrs using the {\sc gadget-2} code \citep{2005MNRAS.364.1105S}.  The best fitting snapshot is found by comparing the bar properties and velocity maps of the 
simulation to observations of the galaxy..

We applied this algorithm to NGC 3621, 1530, and 1300.  For NGC 3621 we found that our model is 
consistent with the ISO and NFW results from \citet{de-Blok:2008oq} and the Einasto model from \citet{2011AJ....142..109C}. It also reproduces the overall shape of the observed RC especially in the inner 5 kpc of the RC. Our result showed that the peak of the parameter PDFs does not always reproduce the actual galaxy. This is why simulations are necessary in addition to the Bayesian analysis. The Bayesian analysis allows us to find plausible axisymmetric models to use for the simulation initial conditions.  But it is the simulations that tell us whether those models will produce bars and 2D velocity maps like those observed.

We were able to reproduce the bar strength from \citet{1998AJ....116.2136A} and the  
kinematic structure of NGC 1530. Our result is comparable to the DiskFit analysis despite  DiskFit being designed to reproduce the velocity map of galaxies with intermediate bar orientation, and,  our azimuthally averaged velocities as well as the radial and tangential velocity moments agree with the results from DiskFit.

Since NGC 1300 is a strongly barred galaxy where the bar is parallel to its position angle, algorithms like DiskFit cannot recover the RC from the velocity map \citep{2010MNRAS.404.1733S,2016A&A...594A..86R}.  Similarly, ROTCUR underestimates the RC for galaxies with such a bar orientation \citep{2015MNRAS.454.3743R}.
Our method is not subject to these same restrictions. Our results show that the mass profile derived from ROTCUR largely underestimates the dynamical mass. This shows the importance of correcting the RC of barred galaxies before using them for mass models analysis.   


This method, while successful in modeling NGC 1300, is not meant to be a replacement algorithm for DiskFit.  Rather, it is a complementary approach that can be applied to galaxies where algorithms such as DiskFit fails.  Based on the results of \citet{2016A&A...594A..86R}, this occurs when bars are within $\sim$10 degrees of the major/minor axes.  Given the frequency of barred galaxies and assuming random orientations, these limits imply that there are many galaxies that would benefit from such an analysis.

In its current state, this algorithm produces viable mass models for specific galaxies.  To get the 'true' mass distribution, we must extend this algorithm to a representative grid of simulations rather than selecting three interesting models.  Such a grid will allow for a test of the uniqueness of our solutions.  Related to this issue is the impact of the inferred stellar parameters on the results.  While our mass-to-light ratios are consistent with those assuming Saltpeter (NGC 1530) and Kroupa (NGC 1300) IMFs, in the future we wish to include these directly into our modeling.  We also wish to extend the method to utilize fits to data cubes rather than velocity maps.


\section*{Acknowledgments}
We thank the anonymous referee for constructive comments.
CC's work is based upon research supported by the South African Research Chairs Initiative (SARChI) of the Department of Science and Technology (DST), the SKA SA and the National Research Foundation (NRF). ND and THR's work were supported by a SARChI's South African SKA Fellowship. LMW is supported by the Natural Sciences and Engineering Research Council of Canada through Discovery Grants
\bibliographystyle{aa}

\appendix
\section{PDF of the model parameters}
\begin{figure*}
 \begin{center}$
  \begin{array}{cc}
   \includegraphics[width=190mm]{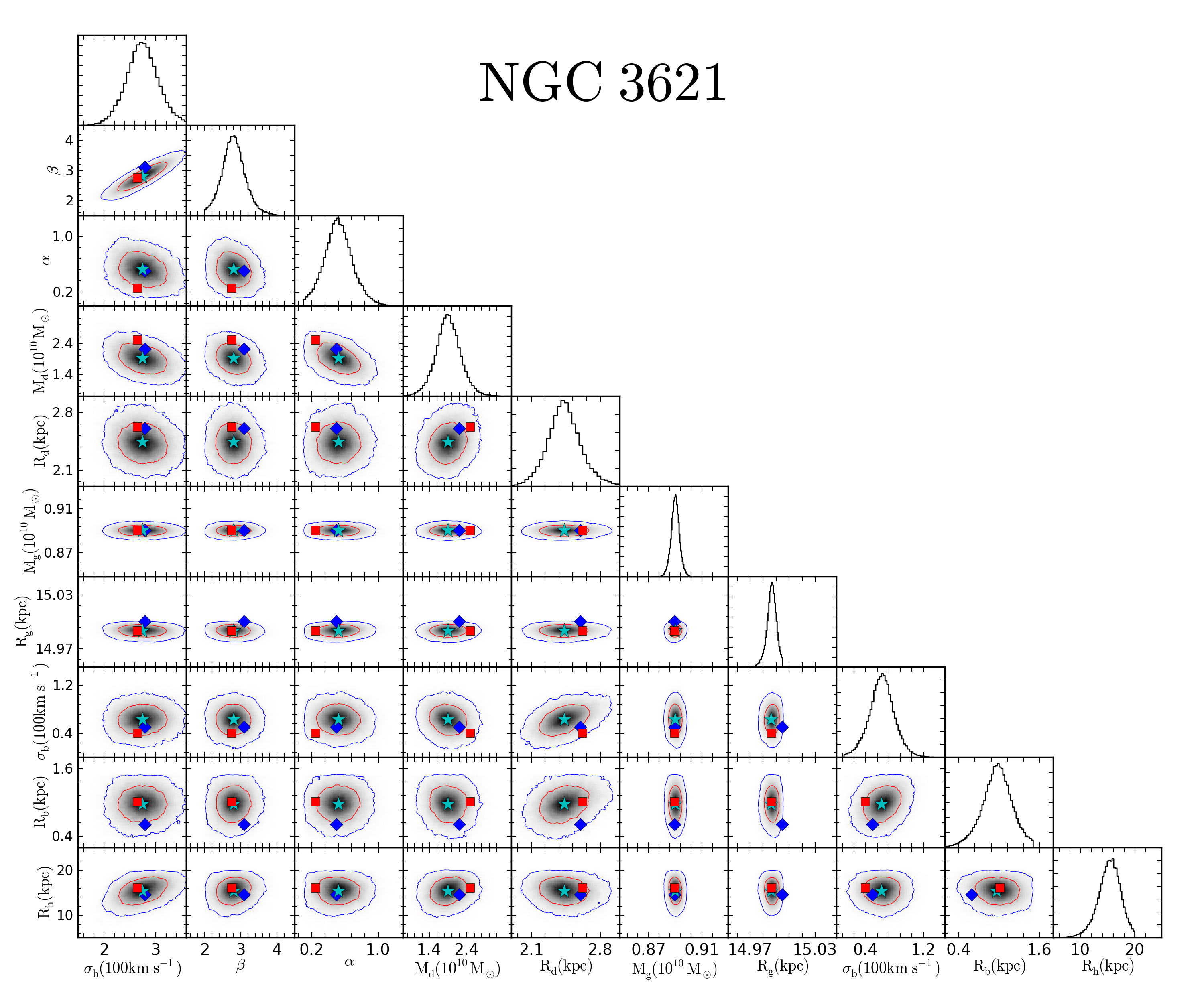}
        \end{array}$
    \end{center}
  \caption{Two-dimensional and one-dimensional PDFs of the model parameters for NGC 3621: the two-dimensional PDFs are shown on the lower triangular of the matrix and the one-dimensional PDFs on the principal diagonals. The red and blue contour in each panel delineates the 68\% and 95\% confidence levels. The cyan stars correspond to Model-C, the blue diamonds to Model-A and the red square to Model-B.}
 \label{figA1}
\end{figure*}

\begin{figure*}
 \begin{center}$
  \begin{array}{cc}
   \includegraphics[width=190mm]{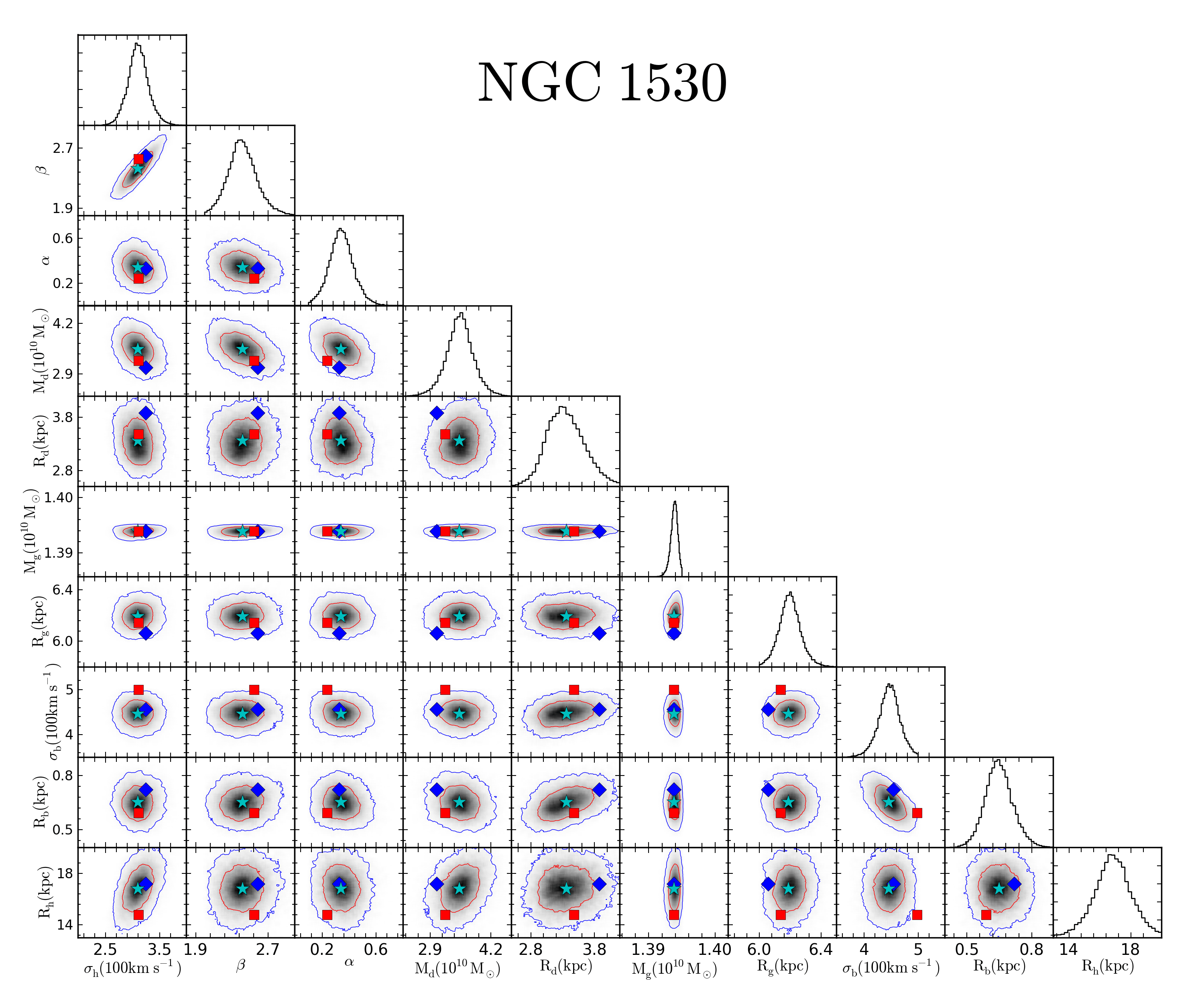}
        \end{array}$
    \end{center}
  \caption{Two-dimensional and one-dimensional PDFs of the model parameters for NGC 1530. 
  Lines and symbols are the same as in Fig. \ref{figA1}.}
 \label{figA2}
\end{figure*}

\begin{figure*}
 \begin{center}$
  \begin{array}{cc}
   \includegraphics[width=190mm]{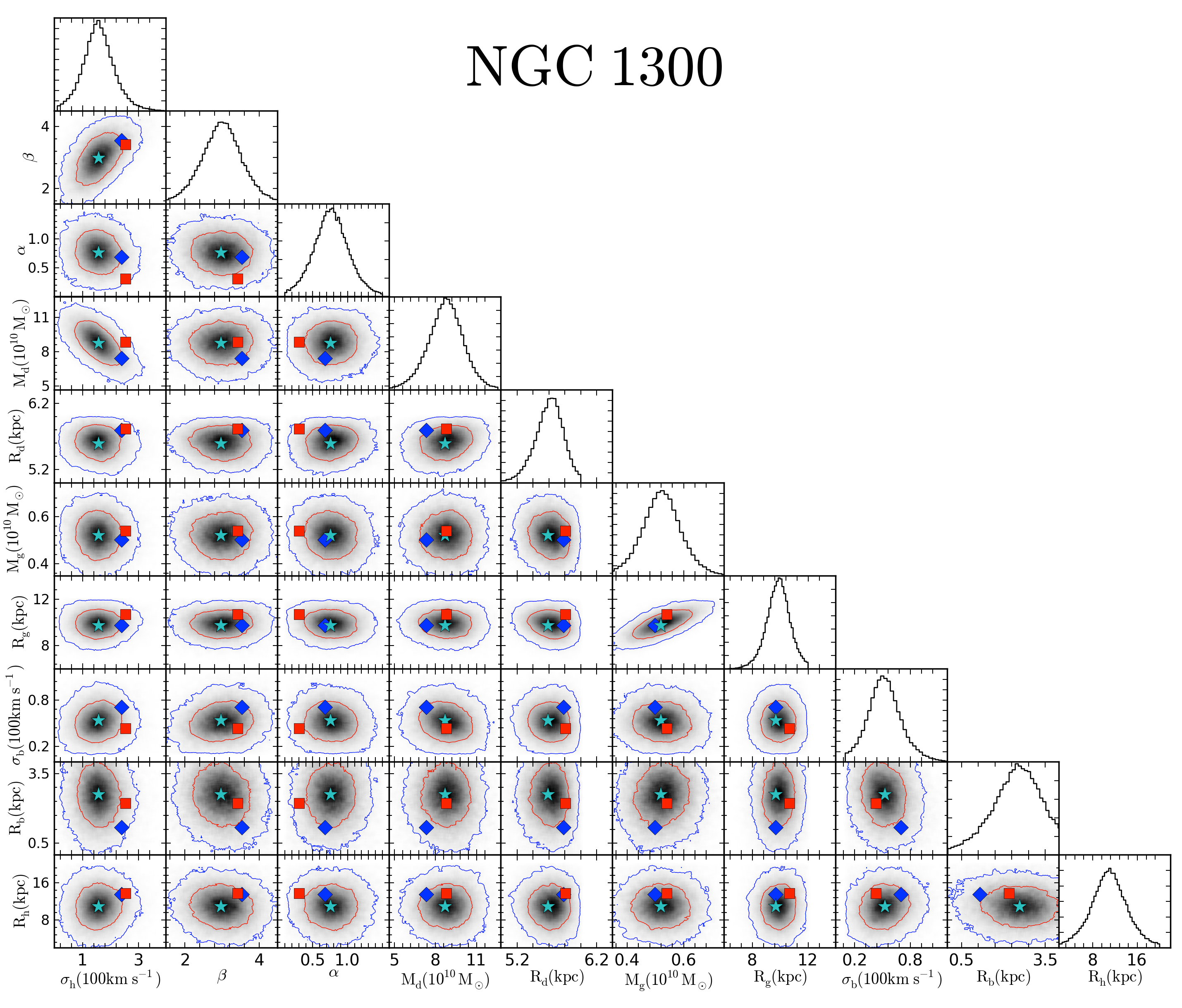}
        \end{array}$
    \end{center}
  \caption{Two-dimensional and one-dimensional PDFs of the model parameters for NGC 1300. Lines and symbols are the same as in Fig. \ref{figA1}}
 \label{figA3}
\end{figure*}

The one and two dimensional PDFs for nine of the free parameters are shown in Figs. \ref{figA1}, \ref{figA2} and \ref{figA3} for NGC 3621, NGC 1530 and NGC 1300 respectively.  We do not show the scale heights for the stellar and gas discs, as the stellar scale height is poorly constrained, and the GalactICS gas disc has a varying scale height determined by the kinematic gas temperature. 
These figures show that the gas parameters are more tightly constrained than the bulge, disc
, and halo parameters. There are a number of correlations apparent between the parameters.  Galaxies with larger sigma prefer smaller alpha and larger beta.  Additionally, there is a correlation between $\sigma_{h}$ and M$_{d}$ that shows the classic disc
-halo degeneracy.

\end{document}